\documentclass[sigconf,nonacm]{acmart}

\AtBeginDocument{%
  }

\usepackage{algorithmic}
\usepackage{graphicx}
\usepackage{amsmath}
\usepackage{textcomp}
\usepackage{xspace}
\usepackage{xcolor}

\usepackage{booktabs}
\usepackage{multirow}
\usepackage{algorithm}
\usepackage{subcaption}
\usepackage{pifont}

\newcommand{\cmark}{\ding{51}}

\graphicspath{ {./figures/} }

\begin{document}

\title{Watch and Crack: Password Inference from Smart-Glasses Video}

\author{Yoav Orenbach}
\orcid{0009-0006-1809-5871}
\affiliation{
  \department{School of Electrical Engineering}
  \institution{Tel Aviv University}
  \city{Ramat Aviv}
  \country{Israel}
}
\email{orenbach@mail.tau.ac.il}

\author{Avishai Wool}
\orcid{0000-0002-8371-4759}
\affiliation{
  \department{School of Electrical Engineering}
  \institution{Tel Aviv University}
  \city{Ramat Aviv}
  \country{Israel}
}
\email{yash@eng.tau.ac.il}

\begin{abstract}
  Typing passwords on smartphones in public places exposes users to video-based side-channel attacks, in which an adversary records the typing session and reconstructs the entered password by analyzing finger movements. Prior video-based keystroke inference attacks have targeted free-form text on tablets, numeric PINs, pattern locks, and passwords under unrealistically simplified conditions.

In this paper we present the first general video-based keystroke inference attack pipeline that recovers rule-based alphanumeric passwords from smartphone QWERTY keyboards, covering all four keyboard layouts, and requiring no assumptions about the victim or their device. Our pipeline uses the built-in camera of the popular Meta Ray-Ban smart glasses to record a victim typing a password on a smartphone in a public setting. It tracks the device and typing fingertips at sub-pixel resolution, predicts keystrokes via a self-supervised ensemble of neural networks, dynamically estimates the on-screen keyboard layout, and computes per-key probability distributions. These video-derived probabilities are then combined with prior password-typing distributions based on 2.19~billion leaked passwords to estimate the rank, and therefore the crack time, of the observed password.
 
We evaluate our pipeline via a user study, against passwords that adhere to a typical enterprise password policy, 
including both human-chosen and password-manager-chosen random passwords. Users typed naturally, some using two thumbs and some using the index finger. 
With our video side channel, human-chosen passwords of up to 16~characters can be cracked in hours to days, and password-manager-chosen random passwords of up to 13~characters can be cracked in under an hour on a modern GPU, reducing password entropy by up to 60~bits relative to the prior baseline. For human-chosen passwords, combining video evidence with prior statistics yields an additional statistically significant reduction of approximately 20~bits compared to either source alone. The attack succeeds at distances up to 1.8\,m, across diverse attacker-victim postures (seated or standing) and viewing angles (from frontal to side-profile), and generalizes across three popular smartphones (Google Pixel~10, iPhone~16, and Samsung Galaxy~A53). 
\end{abstract}



\maketitle
\pagestyle{plain}

\renewcommand{\thefootnote}{}
\footnotetext{This is the full version of a paper accepted to the 2026 ACM
Conference on Computer and Communications Security (CCS). It
contains appendices that are omitted from the conference version.}
\renewcommand{\thefootnote}{\arabic{footnote}}

\section{Introduction}\label{sec:intro}

\begin{figure*}[t]
    \centering
    \begin{subfigure}[b]{0.32\textwidth}
        \centering
        \includegraphics[width=\textwidth]{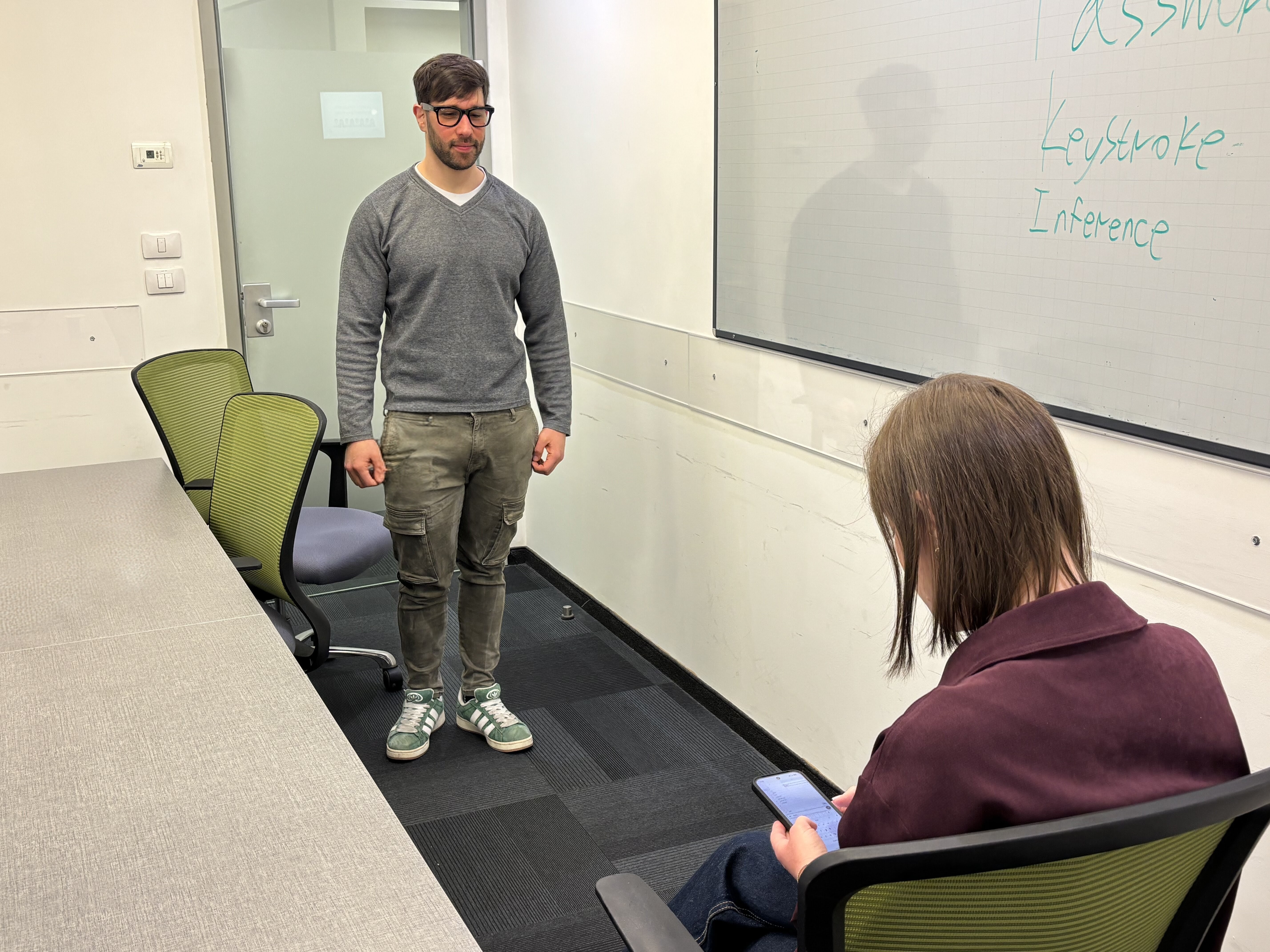}
    \end{subfigure}%
    \hfill
    \begin{subfigure}[b]{0.32\textwidth}
        \centering
        \includegraphics[width=\textwidth]{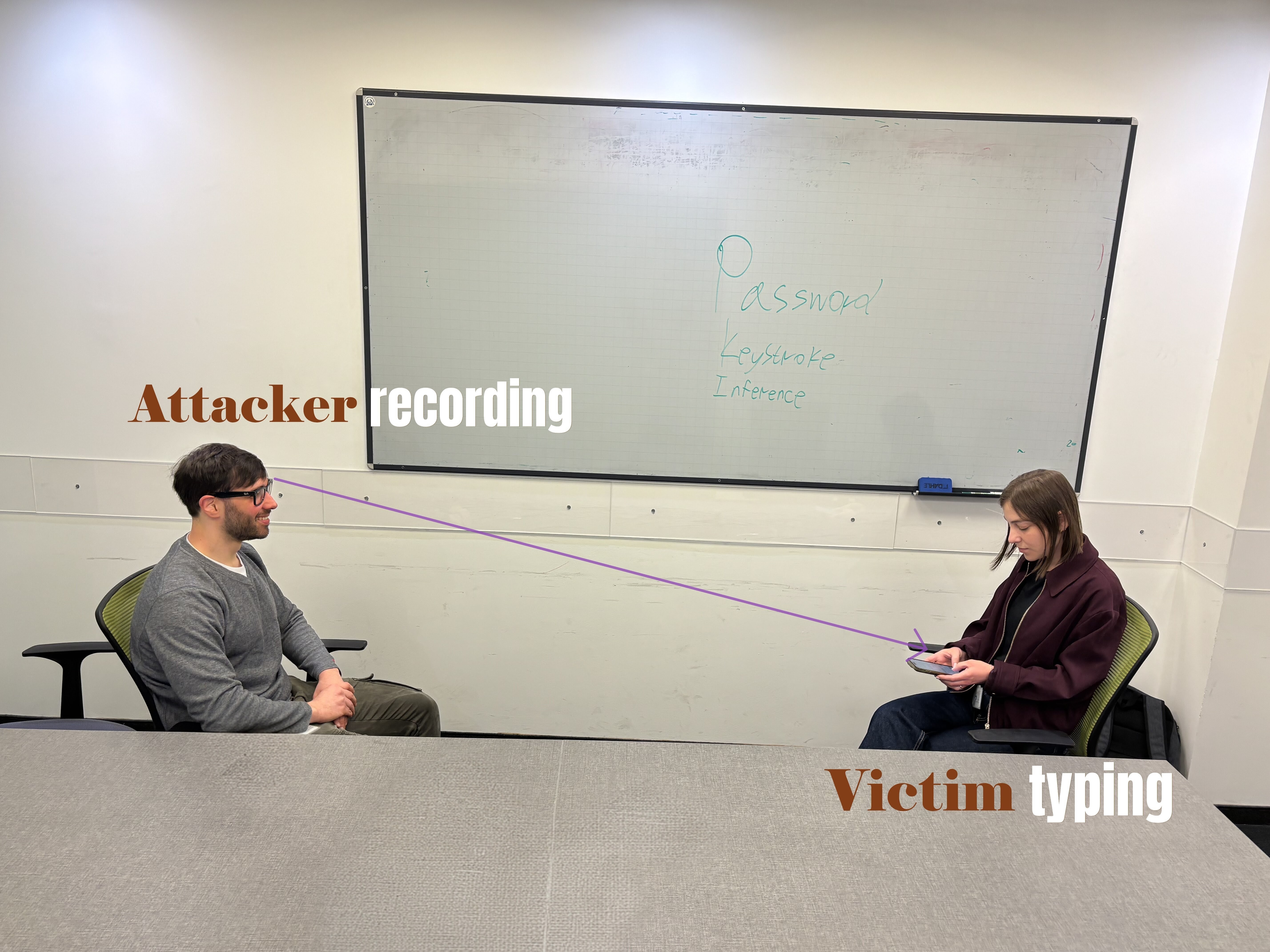}
    \end{subfigure}%
    \hfill
    \begin{subfigure}[b]{0.32\textwidth}
        \centering
        \includegraphics[width=\textwidth]{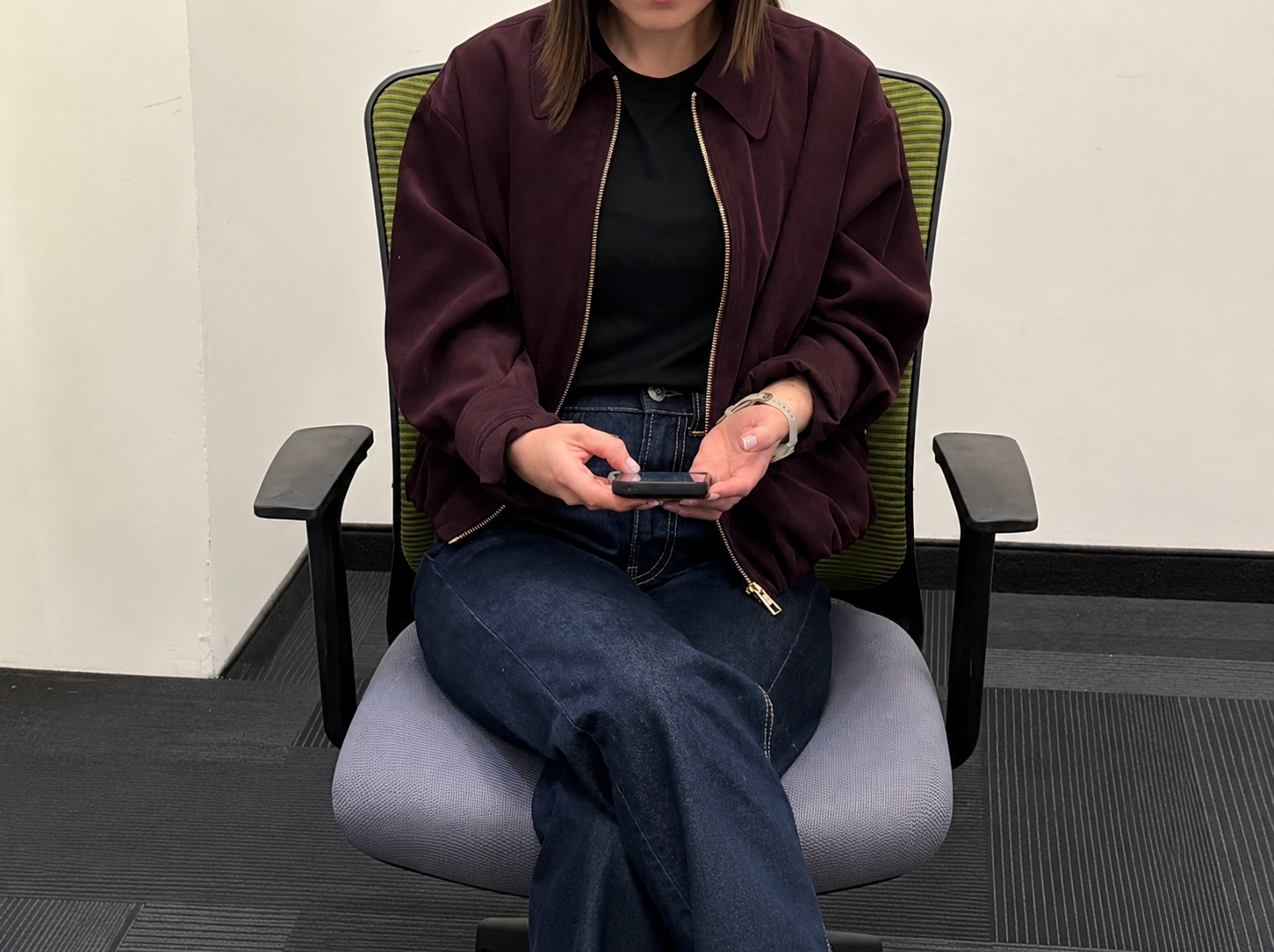}
    \end{subfigure}
    
    \caption{Attack overview. An adversary wearing commercially available Meta Ray-Ban smart glasses records a victim typing a password on a smartphone in a public setting. The attacker later processes the captured video offline to reconstruct the typed password. The attacker may stand or sit across the victim, while having a different view of the typing fingers, highlighting the opportunistic nature of the attack. The rightmost panel shows a cropped frame captured from the glasses themselves.}
    \label{fig:attack_overview}
\end{figure*}

Consider the following scenario: a user sits in a coffee shop, pulls out a smartphone, and logs in by typing a password. The user may be aware of the risks of shoulder-surfing and may even employ a privacy screen protector or the latest privacy display settings. However, an adversary wearing ordinary-looking smart glasses, such as Meta Ray-Ban glasses~\cite{meta-rayban}, could silently record the typing session from a frontal or side view and later reconstruct the password offline (Figure~\ref{fig:attack_overview}).

Machine-learning techniques have become increasingly capable of extracting keystrokes from a variety of side channels, including acoustics~\cite{asonov2004keyboard,zhuang2005keyboard,bwy06}, electromagnetic radiation~\cite{jin2021periscope}, Wi-Fi channel state information~\cite{li2016wifi,chen2018wifi}, and wearable inertial sensors~\cite{wang2015mole,wang2016wearable}. Among these, video-based inference is particularly appealing to opportunistic attackers because it requires no proximity to, or compromise of, the victim's device. Yang~et~al.~\cite{yang2023general} demonstrated the first general video-based keystroke inference attack on free-form text, but their work was limited to tablets where keys are centimeters apart, and to a restricted character set of 26~English letters plus a few punctuation marks. Other video attacks have targeted numeric PINs~\cite{yue2014blind} or pattern locks~\cite{ye2017cracking}, which lack the spatial density and state complexity of a full QWERTY keyboard. 

Video-based attacks on passwords themselves have been attempted, but only under unrealistically simplified conditions. Balagani~et~al.~\cite{balagani2019pilot} recovered short alphanumeric passwords from obfuscated typing videos, but their evaluation targeted computers and ATMs rather than smartphones, and their character set was restricted to lowercase letters and digits. Shukla and Phoha~\cite{shukla2019stealing} attacked passwords by analyzing hand movements on smartphones, but their method was restricted to single-finger typing, did not cover the full QWERTY character set, and did not enforce a strong password composition policy.

To the best of our knowledge, no prior work has demonstrated a general video-based attack that can infer rule-based alphanumeric passwords typed on smartphones, where keys are millimeters apart and all keyboard layouts must be considered for special symbols, without assumptions about the target user or device. This paper fills that gap and shows that combining video-derived keystroke probabilities with publicly available password statistics can reduce password entropy by up to 60~bits, turning ostensibly secure passwords into crackable ones.

\smallskip\noindent\textbf{Why smart glasses?}
Our pipeline is not tied to any particular capture device. It consumes a video of the typing hand and the phone, and applies equally to a concealed smartphone camera or a surveillance camera (\S\ref{sec:threat}). We focus on smart glasses for three reasons: (i) They are a widely deployed consumer camera that yields a stable, hands-free, first-person view of a bystander's phone without a conspicuous posture; (ii) Their prevalence makes a wearer unremarkable in exactly the settings the attack targets~\cite{cnbc_meta_sales_2026}; (iii) Most importantly, the defenses users actually deploy against shoulder-surfing, namely privacy screen protectors and privacy display modes, defeat an adversary who is positioned to read the \emph{screen} but not one who reads \emph{finger movements}.

\smallskip\noindent\textbf{Contributions.} We make the following contributions:

\begin{enumerate}
    \item \textbf{A complete attack pipeline.} We present an eight-step pipeline (\S\ref{sec:pipeline}) that, starting from a single video captured by smart glasses, segments the smartphone screen, tracks fingertips at sub-pixel precision using concept-aware segmentation~\cite{sam3}, estimates keystrokes via a self-supervised ensemble of 3D convolutional networks, dynamically estimates the keyboard layout without knowledge of the device model, and produces per-key probability distributions across all four standard QWERTY keyboard layouts (lowercase, uppercase, numbers, and special symbols). We are the first, to our knowledge, to repurpose concept-aware segmentation for sub-pixel fingertip localization, which makes millimeter-scale key discrimination feasible where landmark trackers fail under occlusion. Our dynamic keyboard-extent estimation recovers keyboard geometry from the layout-switch and shift keystrokes themselves, removing prior reliance on known device dimensions~\cite{shukla2019stealing}. Our detector is self-supervised from a single short video bootstrapped using a concealed label scheme (\S\ref{sec:step5}). And our per-key probabilities are computed in closed form as the Gaussian touch mass within each key's box, propagated through a four-state layout machine.

    \item \textbf{Password rank estimation.} We combine video-derived keystroke probabilities with a prior distribution computed from the Weakpass~4 corpus of 2.19 billion leaked passwords~\cite{weakpass4} using the conflation operator~\cite{hill2011conflations,oren2014new}. Such probability distributions allow the attacker to enumerate the passwords in optimal order \cite{veyrat2012optimal} or in near-optimal order \cite{poussier2016simple,dw17}. The ESrank algorithm~\cite{david2022rank} allows us to estimate the enumeration effort and translate the attack output into concrete crack times,  and thus to quantify the strength of the attack.

    \item \textbf{Comprehensive evaluation.} Through IRB-approved studies with 16~participants (\S\ref{sec:userstudy}), we show that human-chosen passwords of up to 16~characters can be cracked in hours to days, and password-manager-chosen random passwords of up to 13~characters in less than an hour, showing a reduction of up to 60~bits in password entropy compared to the prior baseline alone. The conflated rank of combining video and prior data provides a statistically significant advantage of approximately 20~bits over using either source alone for human-chosen passwords, as evidenced by non-overlapping 95\% confidence intervals. We further demonstrate robustness over distances (0.8--3.0\,m), camera elevations (10--90\,cm), viewing angles (0\textdegree--90\textdegree) and three popular smartphone models (Google Pixel~10, iPhone~16, Samsung Galaxy~A53).

    \item \textbf{User-correction handling.} Real password entry frequently involves backspace corrections that alter the typed sequence. Our pipeline detects and removes such corrections from the predicted keystroke sequence, producing a corrected password candidate for rank estimation.
\end{enumerate}

\section{Background and Related Work}\label{sec:related}

We organize prior work along two dimensions: the \emph{side channel} used (video vs.\ non-video) and the \emph{target modality} (free-form text, PINs/patterns, or full passwords).

\subsection{Non-Video Side Channels}\label{sec:nonvideo}

Adversaries have historically exploited a wide range of physical emanations to extract keystroke data. These include acoustic variations~\cite{asonov2004keyboard,zhuang2005keyboard,bwy06,harrison2023acoustic}, electromagnetic (EM) radiation~\cite{jin2021periscope}, and disruptions in ambient Wi-Fi Channel State Information (CSI)~\cite{li2016wifi,chen2018wifi}. Recent acoustic attacks have demonstrated that deep learning models trained on smartphone microphone recordings can classify laptop keystrokes with up to 95\% accuracy~\cite{harrison2023acoustic}, underscoring the growing sophistication of non-video side channels. In addition, inertial measurement units (IMUs) in wearable devices such as smart watches have been exploited to track kinematic hand movements and infer keystrokes~\cite{wang2015mole,wang2016wearable}.

Indirect video channels have also been explored: reconstructing input from reflections on sunglasses~\cite{xu2013reflections}, tracking ocular micro-movements during soft-keyboard use~\cite{chen2018eyetell}, and analyzing backside vibrations of tablets~\cite{sun2016visible}.

Although potent, these attacks impose strong environmental requirements (e.g., reflective surfaces, high-resolution close-up footage of eyes), require specialized hardware, or necessitate local device compromise.

\subsection{Video-Based Side Channels}\label{sec:videobased}

Direct video side-channel attacks span a range of input modalities. Early work targeted highly structured, short authentication sequences: Ye~et~al.~\cite{ye2017cracking} cracked Android pattern locks from video, Shukla~et~al.~\cite{shukla2014beware} showed that PINs could be inferred from hand movements alone without line-of-sight to the screen, and Chen et al.~\cite{chen2018passcode} further explored PIN leakage from videos capturing the side of the victim. The PILOT system~\cite{balagani2019pilot} demonstrated recovery of both PINs and short alphanumeric passwords from obfuscated typing videos on computers and ATMs. More recently, Dai~et~al.~\cite{dai2024armspy} proposed ArmSpy++, which observes arm posture changes from behind the victim to infer PINs, achieving 83.1\% accuracy within three attempts without requiring any view of the hand or screen. While highly effective, these PIN and pattern attacks are limited to small numeric or geometric search spaces that are fundamentally different from the spatial density of a full QWERTY keyboard.

On the text-inference front, Yang~et~al.~\cite{yang2023general} introduced the first general video-based keystroke inference attack on free-form text, using a self-supervised learning approach that avoids the need for labeled data. Sabra~et~al.~\cite{sabra2021zoom} inferred keystrokes from upper-body movements captured during video calls. Password-specific attacks have also been attempted: Yue~et~al.~\cite{yue2014blind,yue2014my} investigated blind key recognition using Google Glass but were constrained to short numeric passcodes, and Shukla and Phoha~\cite{shukla2019stealing} proposed inferring passwords from the spatiotemporal movements of the typing hand without screen visibility. We compare these closely related works in detail in \S\ref{sec:closely_related}.

Concurrently, spatial computing (VR/AR) has emerged as a new attack vector. Luo~et~al.~\cite{luo2022holologger} and Slocum~et~al.~\cite{slocum2023going} targeted head-gaze movements for keystroke inference on mixed-reality headsets, while Yang~et~al.~\cite{yang2024vr} examined whether VR environments protect users from avatar-based keystroke inference. Most recently, Wang~et~al.~\cite{wang2024gazeploit} demonstrated GAZEploit, a gaze-tracking attack on Apple Vision Pro that infers keystrokes from the eye movements of a user's virtual avatar during video calls. These emerging VR/AR attacks highlight a broadening threat surface, though they operate in a fundamentally different modality of using head or eye tracking within a virtual environment, rather than the physical smartphone typing scenario we address.

\subsection{Closely Related Work}\label{sec:closely_related}

Table~\ref{tab:related} positions our work relative to the most closely related video-based attacks. We discuss each in turn, highlighting the assumptions and limitations that our pipeline overcomes.

Yang~et~al.~\cite{yang2023general} targeted free-form text on tablets, where the keys are substantially wider than on smartphones. Smartphone keys are only millimeters apart, so even small errors in finger localization can map to the wrong key. They also restricted their character set to 26~English letters and limited punctuation, and relied on a language model for spell correction to improve output quality. This approach is fundamentally inapplicable to passwords, which have no linguistic structure, and is further limited by the short length of typical passwords (8--16~characters), which precludes clustering-based approaches that require many keystrokes per video to infer structure.

Sabra~et~al.~\cite{sabra2021zoom} inferred keystrokes from upper-body movements during video chat on physical keyboards. Their reliance on a reference database of commonly used passwords resulted in only 18.9\% of password recovery within top-50~predictions, and their approach does not extend to the spatial density of smartphone keyboards.

Yue~et~al.~\cite{yue2014blind}, Dai~et~al.~\cite{dai2024armspy}, and Chen~et~al.~\cite{chen2018passcode} all target numeric PINs, a search space orders of magnitude smaller than the full QWERTY character set, and they did not address the layout-switching complexity inherent in alphanumeric password entry. Balagani~et~al.~\cite{balagani2019pilot} (PILOT) additionally recovered 8-character alphanumeric passwords from obfuscated typing videos, but their password character set was limited to lowercase letters and digits, with no composition rules requiring uppercase or special characters, and they targeted computers and ATMs rather than smartphones.

The most closely related work to ours is that of Shukla and Phoha~\cite{shukla2019stealing}, who proposed inferring passwords from the spatiotemporal movements of the typing hand using video clips in which the screen is not visible. While their threat model is compelling, their pipeline imposes several restrictive assumptions. First, their work is restricted to \emph{single-finger typing}. Second, their character set covers approximately 62~characters (alphanumeric), excluding most special symbols from the fourth keyboard layout. Third, they let users choose passwords from a leaked corpus with a minimum length of 4~characters and no specific composition requirements for uppercase letters, digits, or special characters, and they did not use randomly generated passwords. Finally, their pipeline requires \emph{manual initialization}: the adversary must manually select anchor points on the hand and phone using the TLD tracker~\cite{kalal2012tld}, and their spatial mapping crucially relies on the \emph{known physical dimensions} of the target device to compute a magnification ratio for mapping pixel movements to key clusters.

In contrast, our work requires no manual initialization, makes no assumptions about device dimensions, supports natural multi-finger typing, covers all keys across all four QWERTY layouts, succeeds against a strict password policy, and quantifies attack strength via formal rank estimation.

\begin{table*}[t]
\centering
\small
\renewcommand{\arraystretch}{1.15}
\begin{tabular}{@{}p{3cm}ccp{3cm}p{3cm}p{3cm}@{}}
\toprule
\textbf{Attack} & \textbf{Target} & \textbf{Device} & 
\textbf{Char.\ Set} & \textbf{Typing Style} & 
\textbf{Password Policy} \\
\midrule
Yang et al.~\cite{yang2023general} & Text & Tablet & Lowercase, space, comma, period, backspace & Natural & N/A (free text) \\
\addlinespace
Shukla \& Phoha~\cite{shukla2019stealing} & Password & Phone & Most QWERTY chars ($\sim$62), excluded most special symbols & Single finger & Min.\ 4 chars, no composition rules \\
\addlinespace
Sabra et al.~\cite{sabra2021zoom} & Password & Physical KB & Full keyboard & Natural & None (reference database) \\
\addlinespace
Balagani et al.~\cite{balagani2019pilot} & Password/PIN & Computer/ATM & Lowercase characters and digits & Natural & 8-char passwords (lowercase + digits); 4-digit PIN \\
\addlinespace
Yue et al.~\cite{yue2014blind} & PIN & Phone/Tablet & Digits (0--9) & Natural & 4-digit PIN \\
\addlinespace
Dai et al.~\cite{dai2024armspy} & PIN & ATM/POS & Digits (0--9) & Natural & 4--6 digit PIN \\
\addlinespace
Chen et al.~\cite{chen2018passcode} & PIN & Phone & Digits (0--9) & Natural & 4-digit PIN \\
\midrule
\textbf{This work} & \textbf{Password} & \textbf{Phone} & \textbf{Full QWERTY, all 4 layouts (115 keys for Gboard~\cite{gboard})} & \textbf{Natural (thumbs or index)} & \textbf{Min.\ 8 chars, at least 1 upper \& 1 digit/symbol} \\
\bottomrule
\end{tabular}
\caption{Comparison with closely related video-based attacks.}\label{tab:related}
\end{table*}

\section{Preliminaries}\label{sec:preliminaries}

Our pipeline relies on several publicly available tools and models. We summarize them briefly.

\smallskip\noindent\textbf{Meta Ray-Ban Smart Glasses.}
Meta Ray-Ban glasses~\cite{meta-rayban} are commercially available smart glasses that resemble conventional eyewear. They embed a forward-facing camera capable of recording video at up to 1080p resolution. With more than 7~million units sold in 2025~\cite{cnbc_meta_sales_2026}, they are increasingly common in public and represent a realistic capture device for an opportunistic attacker. By default, Meta Ray-Ban glasses feature a prominent white \emph{capture LED} that blinks during recording to notify bystanders. In 2026 Meta introduced features so that covering or disabling it automatically disables the camera~\cite{meta_capture_led}---these features were not implemented in the glasses we used in our experiments and we have not evaluated their effectiveness. In our work, we used only the default built-in camera with no optical zoom or external accessories, and we did not hide the LED while recording.

\smallskip\noindent\textbf{MediaPipe Hands.}
MediaPipe Hands~\cite{mediapipe} is a real-time hand-tracking framework developed by Google. It employs a two-stage pipeline consisting of a palm detection model and a hand landmark model, producing 2.5D coordinates (horizontal, vertical, and depth relative to the wrist) for 21~joints per detected hand. While MediaPipe achieves high accuracy in unoccluded scenarios, it is prone to failures when the palm is partially hidden. This is a common occurrence during smartphone typing, where the hand partially encloses the device. These failures manifest themselves as missed detections, hand-identity ambiguity (confusing left and right hands), and fingertip jitter. Our pipeline uses MediaPipe as an initial tracking source and augments it with additional models to overcome these limitations.

\smallskip\noindent\textbf{SAM\,3 (Segment Anything with Concepts).}
SAM\,3~\cite{sam3} is a concept-aware segmentation model that extends the Segment Anything architecture with the ability to segment objects from natural-language text prompts rather than requiring bounding-box or point inputs. Given a text prompt such as \emph{``cell phone''} or \emph{``fingers,''} SAM\,3 produces pixel-level segmentation masks for all matching instances in the image. We leverage SAM\,3 for two purposes: (1)~segmenting the smartphone screen to track its position and corners across frames, and (2)~segmenting typing fingertips to achieve sub-pixel localization that compensates for MediaPipe's tracking errors.

\smallskip\noindent\textbf{Optical Flow.}
The Horn--Schunck optical flow algorithm~\cite{horn1981optical} estimates per-pixel motion vectors between consecutive video frames by jointly minimizing a brightness-constancy term and a spatial-smoothness regularizer. We use optical flow to propagate fingertip positions across frames in which MediaPipe fails to produce valid detections, bridging temporal gaps in the initial hand-tracking.

\smallskip\noindent\textbf{R3D-18.}
R3D-18~\cite{tran2018closer} is a 3D residual convolutional neural network designed for video action recognition, pretrained on the Kinetics-400 human action dataset. We fine-tune it on a 16-frame temporal window to classify whether a fingertip is performing a keypress, and we train an ensemble of five such networks to improve robustness.

\smallskip\noindent\textbf{Conflation.}\label{sec:conflation}
When an attacker obtains multiple independent probability distributions over the same discrete variable, for instance, one from video analysis and another from prior password statistics, a principled method is needed to combine them. Hill~\cite{hill2011conflations} introduced \emph{conflation} as a general method to reconcile multiple probability distributions. Given two distributions $P$ and $Q$ over the same finite set $\{V_1, \dots, V_N\}$, their conflation is defined as the normalized element-wise product:
\begin{equation}
    \hat{P}(V_i) = \frac{P(V_i) \cdot Q(V_i)}{\sum_{j=1}^{N} P(V_j) \cdot Q(V_j)}
    \label{eq:conflation}
\end{equation}Hill showed that conflation is the unique distribution that minimizes the total loss of Shannon information relative to the input distributions and that it automatically assigns more weight to distributions with smaller variance (i.e., more confident measurements).

Conflation has been successfully applied in the side-channel cryptanalysis literature. Oren, Weisse, and Wool~\cite{oren2014new} used conflation as the core probabilistic reconciliation mechanism in a constraint-based template attack on AES, where multiple noisy side-channel measurements of the same key byte are combined into a single high-confidence distribution. In our setting, we apply conflation analogously: for each keystroke position, we conflate the video-derived probability distribution (from Step~6 of our pipeline) with the a-priori password-typing distribution (from \S\ref{sec:prior}) to produce a \emph{conflated} distribution that leverages both the visual evidence and the statistical regularities of human password composition.

\smallskip\noindent\textbf{ESrank (Exponential Sampling Rank Estimation).}
When a side-channel attack produces a probability distribution over candidate keystrokes, the adversary must enumerate passwords in descending order of likelihood \cite{veyrat2012optimal}. Computing the exact rank of the correct password, namely its position in this sorted list is computationally prohibitive for large keyspaces via na\"ive brute-force methods. A significant advance in this domain was the ESrank algorithm~\cite{david2022rank}, which provides poly-logarithmic rank estimation.
The ESrank algorithm efficiently estimates the rank of a given key within a combinatorially large keyspace without explicit enumeration. Formally, given $d$~independent subkey spaces with probability distributions $P_1, \dots, P_d$ sorted in decreasing order, and a target key $k^* = (k_1, \dots, k_d)$ with probability $p^* = \prod_{i=1}^{d} P_i(k_i)$, ESrank estimates the number of keys whose probability exceeds~$p^*$, that is, the position of~$k^*$ in the probability-sorted list of all $\prod_{i=1}^{d} n_i$ possible keys. The algorithm runs in poly-logarithmic time relative to the keyspace size, making it practical for the large spaces encountered in password cracking. 

In the password-cracking domain, a rank of~$r$ means that exactly $r-1$ password candidates are assigned a higher probability than the true password, so an adversary enumerating in decreasing probability order reaches it after at most $r-1$ incorrect guesses. We report $\log_2 r$ bits, so $20$~bits denotes an enumeration of $2^{20}$~candidates, and a $60$-bit reduction denotes a search space smaller by a factor of $2^{60}$. We use ESrank as the final stage of our pipeline to convert per-key probability distributions into a concrete password rank and, by extension, an estimated crack time.

\section{Threat Model}\label{sec:threat}

We target a realistic scenario in which users type passwords on smartphones in common public settings such as cafés, airport lounges, or study halls.

\smallskip\noindent\textbf{Attacker capabilities.} The attacker wears commercially available smart glasses (e.g., Meta Ray-Ban~\cite{meta-rayban}, see \S\ref{sec:preliminaries}) and records the victim from a frontal or oblique viewing angle using the glasses' built-in camera without zoom or external equipment. The attacker then processes the recorded video offline using publicly available tools described in \S\ref{sec:preliminaries}: MediaPipe~\cite{mediapipe} for initial hand tracking, SAM\,3~\cite{sam3} for device and finger segmentation, and R3D-18~\cite{tran2018closer} for keystroke classification. The attacker knows that the victim uses one of the commonly deployed QWERTY keyboard applications (e.g., Gboard~\cite{gboard}, the iOS keyboard~\cite{apple_ios_keyboard}, or the Samsung keyboard~\cite{samsung_keyboard}), although the attacker does not need to know which one is installed.

Consistent with a general, assumption-free attack, the attacker does \emph{not} require:
\begin{itemize}
    \item The attacker has \emph{no prior knowledge} of the victim's specific smartphone model, screen dimensions, or which keyboard application is installed. The attacker compensates for the unknown keyboard configuration by running the probability-estimation stage for each candidate layout in parallel, as described in \S\ref{sec:step6}.
    \item The attacker has \emph{no labeled data} from, or prior observations of the victim, hence the password may be stolen opportunistically.
    \item The attacker has \emph{no access to} any sensor, device, or communication channel beyond the glasses' camera.
    \item The victim's screen may not be directly readable (e.g., due to a privacy screen protector or privacy display mode).
\end{itemize}

\smallskip\noindent\textbf{Offline guessing model.}
We assume the adversary has already obtained the target's password \emph{hash}, for example from a breached credential database. Our attack outputs a probability-ranked list of candidate passwords, and this assumption is what allows those candidates to be tested offline at the guessing rate used in \S\ref{sec:step8}. Salting does not weaken the model, since the salt is stored alongside the hash and disclosed by the same breach. The assumed crack rate corresponds to fast hash functions, and \S\ref{sec:step8} explains how our reported times rescale under a slow key-derivation function. Mitigations such as attempt limiting, lockout, and two-factor authentication only restrict online login attempts rather than offline guessing, and so lie outside this model by construction (\S\ref{sec:defenses}).

\smallskip\noindent\textbf{Identifying a password-entry session.}
Our pipeline analyzes a video it is given, and it neither classifies sessions nor reads screen \emph{content}. The attack is opportunistic, and the adversary records whenever there is reason to believe a password is being entered, hence identifying when a password is being entered is largely out of our research scope. Nevertheless, we note several cues can that could assist the adversary in identifying password entry without requiring a legible screen---other cues may well be possible. The most direct cue is the privacy display mode noted above. Where a device offers such a mode and the user enables it for password entry, its activation visibly alters the appearance of the screen while leaving the content unreadable, so the very defense that motivates our attack also marks the moment worth recording. Other possible cues can come from off-camera sources such as a compromised component of the breached service from which the hash was stolen, or a collaborator who observes an authentication attempt in progress. An adversary willing to act rather than only observe can prompt a session in conversation, for instance by remarking on a stock price so that the victim opens a portfolio application. A false trigger costs only wasted offline computation and does not affect the results we report for genuine sessions.

\smallskip\noindent\textbf{Capture device.}
No stage of the pipeline depends on the video originating from smart glasses. The attack only requires a clear, stable view of the typing hand and the phone and applies unchanged to a concealed smartphone camera, a fixed surveillance camera, or a body-worn action camera of comparable quality. Section~\ref{sec:scenarios} characterizes the capture regime in which it succeeds in terms of distance, elevation, and angle rather than any glasses-specific property.

\smallskip\noindent\textbf{Password policy.} In our testing, the passwords adhere to a realistic and widely deployed password composition policy that requires a minimum of eight characters with at least one uppercase letter and at least one digit or special character. Although recent NIST guidelines (SP~800-63B) de-emphasize composition rules in favor of length, such traditional policies remain pervasive in deployed web services and enterprise systems~\cite{blocki2013optimizing,florencio2007large,ur2016perceptions,komanduri2011passwords}.

\section{Properties of Smartphone Password Typing}\label{sec:properties}
Before presenting our pipeline, we describe two properties of smartphone password entry that shape its design.

\subsection{Smartphone Typing Characteristics}\label{sec:typing}

A standard smartphone QWERTY keyboard exposes four layouts (illustrated for Gboard, iOS, and Samsung in Figures~\ref{fig:gboard_layouts}--\ref{fig:samsung_layouts} in Appendix~\ref{app:layouts}): (1)~lowercase letters, (2)~uppercase letters (reached via the shift key~$\Uparrow$), (3)~numbers and common symbols (reached via the layout-switch key \texttt{?123}), and (4)~additional special symbols (reached via a second shift from layout~3). Typing a password adhering to a password policy necessarily involves layout transitions, increasing both the number of keystrokes and the complexity of the inference task. For example, the password \texttt{Myspace0} requires the typed sequence $[\Uparrow, \texttt{M}, \texttt{y}, \texttt{s}, \texttt{p}, \texttt{a}, \texttt{c}, \texttt{e}, \texttt{?123}, \texttt{0}]$. This amounts to ten keystrokes for an eight-character password.

Crucially, different keyboard applications arrange numbers and symbols differently. Gboard and the iOS keyboard share nearly identical lowercase and uppercase layouts but diverge in their numbers and symbols layouts: certain special characters that Gboard places in the numbers layout appear in the fourth layout on iOS, requiring an additional layout switch. Conversely, the Samsung keyboard takes a different approach entirely, presenting a five-row layout with numbers directly above the letter rows, eliminating the need for a layout switch to reach digits. As a result, the same password may require different typed sequences, and therefore different touch counts, depending on the keyboard application. Because in our attack model the attacker does not know the phone brand or keyboard application which the victim uses, the pipeline must account for all common configurations, as detailed in Step~6 (see \S\ref{sec:step6}).

\subsection{A-Priori Password Typing Distribution}\label{sec:prior}

Human-chosen passwords follow well-studied composition patterns~\cite{ur2016perceptions,komanduri2011passwords}. As a result, the \emph{typed sequences}, including layout-switch and shift keystrokes, also exhibit regularity that can serve as prior information for an attacker.

Therefore, in an offline step computed once, we construct a prior distribution from the Weakpass~4 corpus of 2.19 billion leaked passwords~\cite{weakpass4} for each of the three keyboard applications. For each leaked password, we convert it to its full typed sequence starting from the default lowercase layout, determining every required keystroke including layout switches and shifts. For each typed-sequence length~$d$ and each position $i \in \{0, \dots, d{-}1\}$, we compute the empirical frequency of every character across all passwords of that length. This yields per-position probability distributions $P_1^{\text{prior}}, \dots, P_d^{\text{prior}}$ that serve as a baseline: an attacker who knows only the number of keystrokes can already perform a dictionary attack sorted by these prior probabilities.

\section{The Video Processing Pipeline} \label{sec:pipeline}

\begin{figure}[t]
    \centering
    \includegraphics[scale=0.5]{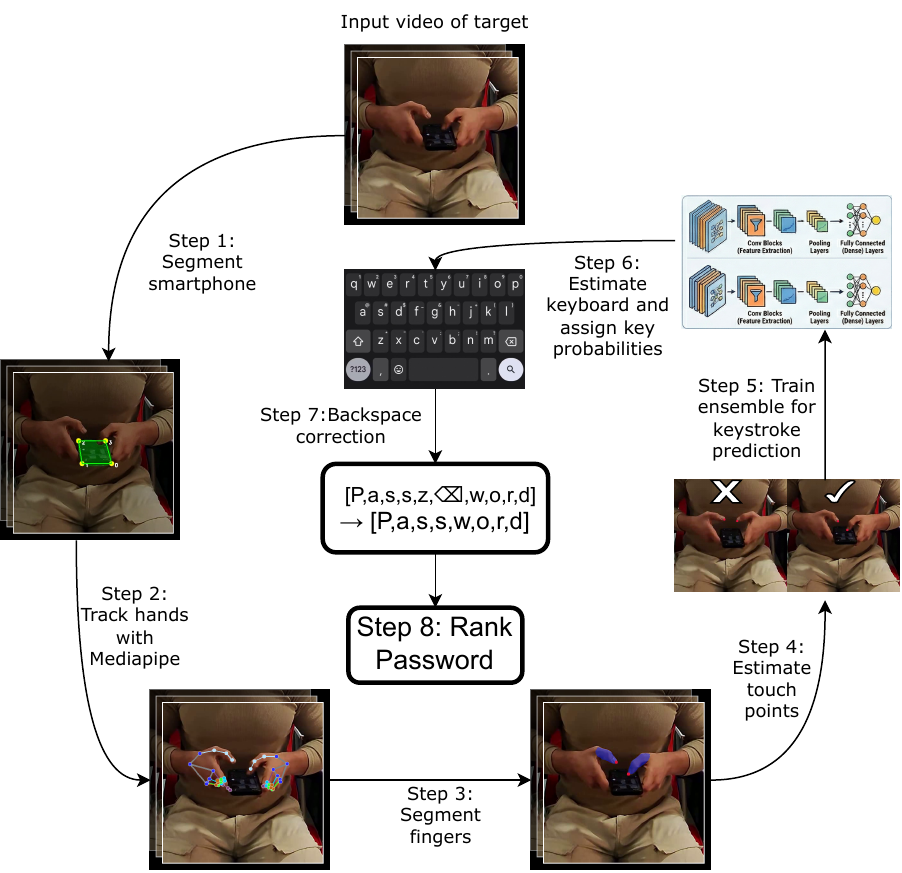}
    \caption{System architecture of the eight-step attack pipeline. 
    \textbf{Step~1}: The smartphone screen is segmented and tracked 
    using SAM\,3. 
    \textbf{Step~2}: Initial fingertip positions are obtained via MediaPipe 
    and optical flow.
    \textbf{Step~3}: SAM\,3 refines finger localization to sub-pixel precision. 
    \textbf{Step~4}: Candidate keypress events are identified from the fingertip signal. 
    \textbf{Step~5}: An ensemble of five R3D-18 networks predicts final keystrokes from the pseudo-labeled data.  
    \textbf{Step~6}: The keyboard layout is dynamically estimated and per-key probabilities. 
    \textbf{Step~7}: Backspace keys are detected and corrected.
    \textbf{Step~8}: The ESrank algorithm estimates the password rank. 
    }
    \label{fig:pipeline}
\end{figure}

This section details our eight-step pipeline, illustrated in Figure~\ref{fig:pipeline}. Given a single video of the victim typing a password, we first perform an automatic preprocessing pass that localizes the hands in the early frames using MediaPipe Hands~\cite{mediapipe_hand_api} and crops the video around a detected anchor point. This yields a tightly framed input around the typing region, which is fed into the following pipeline.

\subsection{Step 1: Smartphone Tracking}\label{sec:step1}

To account for the natural movement of the smartphone during typing, we segment the screen in every frame using the concept-aware segmentation model SAM\,3~\cite{sam3} with the text prompt \emph{``cell phone''}  by utilizing the API provided by Ultralytics~\cite{ultralytics_sam3}. From each segmentation mask, we extract the four screen corners by approximating a polygon, yielding a per-frame quadrilateral that tracks the screen throughout the video.

\subsection{Step 2: Initial Hand Tracking}\label{sec:step2}

We obtain initial fingertip positions using MediaPipe Hands real-time API~\cite{mediapipe_hand_api}. On smartphones, one or both palms are frequently occluded, causing MediaPipe to fail, exhibit hand ambiguity, or produce jittery finger predictions. To bridge gaps, whenever MediaPipe fails to produce valid coordinates for a frame, we propagate the last successful prediction forward using optical flow~\cite{horn1981optical}. If MediaPipe fails on the initial frames, we back-fill with the first valid detection.

\subsection{Step 3: Sub-Pixel Finger Tracking}\label{sec:step3}

Because smartphone keys are millimeters apart, we require sub-pixel finger localization. We again leverage SAM\,3~\cite{sam3}, this time with a single generic concept prompt \emph{``fingers''} to track every possible typing finger. SAM\,3 returns multiple candidate segmentation masks, potentially including non-typing fingers and, crucially, multiple overlapping masks covering the same physical finger (e.g., a partial segment of a thumb alongside a larger segment of the same thumb). We therefore merge overlapping detections, and for each merged mask we extract the fingertip as the point of maximum $y$-coordinate (i.e., the lowest pixel in image space), averaging nearby boundary pixels within 2~pixels of this extremum to suppress tracking noise. We then select the masks whose resulting fingertip points have the smallest Euclidean distance to the corresponding MediaPipe+optical-flow points from Step~2. Finally, we apply a frame-by-frame tracker based on the Hungarian algorithm~\cite{kuhn1955hungarian} to maintain consistent finger identities across frames. The output is, for each of $f$~typing fingers across $n$~frames, a sequence of sub-pixel fingertip coordinates.

\subsection{Step 4: Touch Estimation}\label{sec:step4}

Before analysis, we sanitize the per-finger signals from Step~3 by discarding any finger whose track is missing in more than half of the frames, and linearly interpolating short gaps in the surviving signals to yield a continuous fingertip trajectory per finger. We then analyze the fingertip coordinate signal to identify candidate keypress events. A fingertip coordinate pair $(f_j, \text{frame}_i)$ is considered a candidate keypress only if all of the following conditions hold:

\begin{enumerate}
    \item The vertical ($y$) position exhibits a local peak, with prominence exceeding a data-driven threshold, reflecting the downward press-and-release motion.
    \item The fingertip acceleration exhibits a corresponding negative peak, reflecting the abrupt deceleration when the finger contacts the screen.
    \item The fingertip lies within the screen bounds estimated in Step~1.
\end{enumerate}

To set the prominence threshold in condition~(1), and following the approach of Yang~et~al.~\cite{yang2023general}, we model the distribution of peak prominence values as a two-component Gaussian Mixture Model (GMM) over keypress vs.\ non-keypress peaks. The threshold is set at the minimum between the two component modes, approximating equal false-alarm and misdetection rates. To filter spurious peaks caused by hesitations or subtle movements of non-typing fingers, only peaks exceeding this threshold and meet conditions~(2) and~(3), are retained as high-confidence pseudo-labels for the next step.

\subsection{Step 5: Keystroke Prediction via Ensemble CNNs}\label{sec:step5}

The pseudo-labels from Step~4 serve as training data for an ensemble of five 3D convolutional neural networks. We use R3D-18~\cite{tran2018closer} as the backbone, and replace its original 400-class output layer with a two-layer classification head: a 512$\to$256 bottleneck with ReLU activation, bracketed by 50\% dropout layers, followed by a final linear layer producing binary keypress/non-keypress predictions. The aggressive dropout is critical given that each video yields only a few dozen positive training examples.

\smallskip\noindent\textbf{Dataset construction.} For each finger~$j$ and frame~$i$, we assign a label:
\begin{itemize}
    \item \emph{Positive}: frames within a 1-frame radius of a high-confidence pseudo-label.
    \item \emph{Excluded from training} (test-only): peaks that did not pass the prominence threshold but had nonzero prominence (possible missed touches), along with a 4-frame exclusion zone around them and around the positive labels.
    \item \emph{Negative}: all remaining frames.
\end{itemize}
The excluded set is the mechanism by which the ensemble recovers keystrokes that Step~4 missed. Sub-threshold peaks are genuinely ambiguous. Labeling them positive would inject spurious keystrokes into training, while labeling them negative would teach the network to suppress exactly the low-prominence presses we most want to detect. We therefore withhold them from the loss entirely, train only on high-confidence positives and confident negatives, and at inference score the withheld frames with the rest of the video, unchanged; the 4-frame zones do the same for frames adjacent to a labeled event, whose class is uncertain because the press boundary is not known to frame precision. This train/inference asymmetry is what lets the ensemble outperform, rather than merely reproduce, the procedure that supervised it (\S\ref{sec:ablation}).

Each frame is cropped to a $56 \times 56$~pixel region centered on the fingertip. The network input is a window of 16~consecutive crops. We apply Mixup~\cite{zhang2020mixup} - a data-augmentation technique that generates synthetic training samples by linearly interpolating pairs of inputs and their labels. It has been shown to smooth decision boundaries and reduce overfitting, which is particularly valuable in our setting where the training data is derived from a single short video with limited positive examples.

\smallskip\noindent\textbf{Inference.} After training, all five classifiers scan the full video. We average their probability scores and, rather than selecting a single score threshold, we evaluate multiple thresholds $\{0.2, 0.3, \dots, 0.9\}$ to produce a set of candidate keystroke counts. This is critical because a single missed or spurious keystroke renders the entire password guess incorrect. By maintaining multiple candidate lengths, the attacker can run parallel dictionary attacks, substantially improving the probability that one list has the correct length.

\subsection{Step 6: Video Probability Estimation}\label{sec:step6}

With the predicted keystrokes, screen corners, and fingertip trajectories in hand, we now estimate the probability of every key for each keystroke event.

\smallskip\noindent\textbf{Dynamic keyboard estimation.}
Our approach does not use a-priori knowledge of the phone dimensions or of the keyboard in use. Thus we dynamically estimate the fraction of the screen occupied by the keyboard. We observed that for a rule-based password on a standard 4-row keyboard (e.g., Gboard, iOS), the user must press the layout-switch key \texttt{?123} (located in the last row) at least once to reach digits, and a digit (located in the first row of the numbers layout) at least once. For 5-row keyboards that display digits directly on the primary layout (e.g., Samsung), the layout-switch key may not be pressed. However, the shift key~$\Uparrow$, located in the second-to-last row, must still be pressed at least once to satisfy the uppercase requirement of the password policy. In both cases, the observed touch points span at least two distinct keyboard rows, enabling dynamic estimation of the keyboard extent. We exploit this by computing, for every predicted keystroke, the ratio of the fingertip's vertical position relative to the screen extent:
\begin{equation}
    r_i = \frac{y_i - y_{\text{bottom}}}{y_{\text{top}} - y_{\text{bottom}}}
\end{equation}
where $y_{\text{top}}$ and $y_{\text{bottom}}$ are the $y$-coordinates on the top and bottom screen edges at the fingertip's $x$-coordinate (accounting for perspective). The minimum and maximum ratios over all keystrokes approximate the bottom and top of the keyboard, respectively.
This estimate relies on the observed keystrokes spanning at least two keyboard rows, which the password policy we target guarantees.

\smallskip\noindent\textbf{Per-key probability computation.}\label{sec:per_key_prob}
Given the estimated keyboard corners for each frame, we apply a homography from the estimated corners to a normalized keyboard template. We constructed one such template per keyboard application (Gboard~\cite{gboard}, iOS~\cite{apple_ios_keyboard}, Samsung~\cite{samsung_keyboard}) by encoding the standard key arrangement of each layout in a unit-width coordinate system, with key centers and bounding boxes derived from each application's documented layout geometry. Separate templates are maintained for each of the four layout states. The fingertip position is projected into the same normalized coordinate space. The probability assigned to each key is computed as the integral over each key's bounding box of an axis-aligned Gaussian with standard deviations $\sigma_x$ and $\sigma_y$, centered at the detected touch point $(X, Y)$. Specifically, if a key's bounding box is $[x_0, x_1] \times [y_0, y_1]$ the probability assigned to the key is:
\begin{equation}
    \Pr(\text{key}) =
    \int_{x_0}^{x_1}\int_{y_0}^{y_1}
    \exp\!\left(-\frac{(x - X)^2}{2\sigma_x^2} - \frac{(y - Y)^2}{2\sigma_y^2}\right)
    dy\, dx
    \label{eq:key_integral}
\end{equation}

\noindent Since the Gaussian is separable, Equation~(\ref{eq:key_integral}) admits a closed-form solution via the error function:
\begin{align}
& \Pr(\text{key}) = \nonumber \\
 &\Bigg[
        \mathrm{erf}\!\bigg(\frac{x_1 - X}{\sqrt{2}\,\sigma_x}\bigg) -
        \mathrm{erf}\!\bigg(\frac{x_0 - X}{\sqrt{2}\,\sigma_x}\bigg)
    \Bigg]
    \cdot
    \Bigg[
        \mathrm{erf}\!\bigg(\frac{y_1 - Y}{\sqrt{2}\,\sigma_y}\bigg) -
        \mathrm{erf}\!\bigg(\frac{y_0 - Y}{\sqrt{2}\,\sigma_y}\bigg)
    \Bigg]
    \label{eq:key_erf}
\end{align}

\noindent This formulation assigns each key the total probability mass of the touch distribution that falls within its physical area
and naturally accounts for variable-width keys, such as the space bar, through their larger integration bounds.

The standard deviations $\sigma_x$ and $\sigma_y$ are fixed globally across all keys, reflecting the assumption that touch uncertainty is a property of the user and the device rather than of the individual key being pressed. Specifically, we set $\sigma_x = 0.5 \cdot w_{\min}$ and $\sigma_y = 0.5 \cdot h$, where $w_{\min}$ is the width of a standard alphanumeric key and $h$ is the uniform key height we set in normalized coordinates. This choice ensures that a touch landing at a key boundary lies one standard deviation from the key center, naturally capturing the spatial uncertainty of finger placement. We validate this choice against alternative coefficient values in \S\ref{sec:sigma_choice}.

We account for keyboard state across the password entry sequence: the first keystroke considers only the lowercase QWERTY layout, and the second considers lowercase, uppercase, and numbers. From the third keystroke onward, all four layouts are considered. Thus, from the third keystroke, the value from equation~\ref{eq:key_erf} is assigned to all 4 keys occupying the same position in the 4 layouts. These per-key probabilities are summed across all applicable layouts and normalized to form a valid probability distribution.

\smallskip\noindent\textbf{Multi-configuration execution.}
Because different keyboard applications define different key placements in the numbers and symbols layouts, a single normalized template may not match the victim's actual keyboard. To maintain the assumption-free design, the attacker runs the probability estimation independently for each candidate keyboard configuration (i.e., Gboard, iOS, Samsung). Each configuration maps the same set of predicted touch points to its own normalized layout, producing a separate set of per-key probability distributions. Importantly, because keyboards differ in the number of layout switches required, a password that requires $x$~touch points on one configuration may require $x' \neq x$ on another. This multi-configuration integrates naturally with the existing parallel-attack framework established by the ensemble's variable-length output (Step~5).

\subsection{Step 7: Backspace Detection}\label{sec:step7}

When typing complex passwords, users may make errors and correct them with the backspace key. After computing per-key probabilities (Step~6), if the backspace key appears among the top-5 most probable keys for the $j$-th keystroke ($j > 0$), we remove both keystroke~$j$ (the backspace itself) and keystroke~$j{-}1$ (the erroneously typed character). For $x$~predicted keystrokes with $b$~detected backspaces, this yields a reduced list of $x - 2b$~keystrokes. Both the original and reduced lists are retained for parallel dictionary attacks.

\subsection{Step 8: Rank Estimation}\label{sec:step8}

As the final step, we quantify the attack's effectiveness using password rank estimation.

\smallskip\noindent\textbf{Length validation.} If none of the candidate keystroke counts from the previous steps matches the true typed-sequence length, the attack is declared a failure regardless of key-position accuracy. This is a stricter criterion than reporting per-character accuracy, as used in some prior work.

\smallskip\noindent\textbf{Rank computation.} For a typed sequence of length~$d$, we compute three ranks using the ESrank algorithm~\cite{david2022rank}:
\begin{enumerate}
    \item \emph{Prior rank}: using only the a-priori distribution from \S\ref{sec:prior}.
    \item \emph{Video rank}: using only the video-derived per-key distributions from Step~6.
    \item \emph{Conflated rank}: combining the prior and video distributions via the conflation operator (recall \S\ref{sec:conflation}, Equation~\ref{eq:conflation}), which takes the element-wise product of the per-position probability vectors and renormalizes.
\end{enumerate}
The three keyboard configurations of Step~6 are \emph{not} merged or normalized into a single distribution. Each configuration yields its own typed-sequence hypothesis and its own per-key distributions, and each is carried through Steps~7 and~8 independently to produce a separate ranked candidate list, which the adversary enumerates as a separate dictionary attack. Accordingly, the final rank is multiplied by the total number of parallel dictionary attacks the adversary must execute. This count is the product of three factors: the number of candidate keystroke counts retained from Step~5, the number of keyboard configurations from Step~6 that yield a matching typed-sequence length, and the number of backspace variants from Step~7 (original and reduced lists). This multiplicative penalty ensures that the rank reflects the true cost of the attacker's uncertainty across all three dimensions. The rank~$r$ is converted to entropy bits ($\log_2 r$) and to crack time assuming an adversary speed of $10^{10}$~guesses per second against an offline verifier storing a fast hash (\S\ref{sec:threat}), consistent with contemporary GPU-based cracking~\cite{melicher2016neural,huang2024probhashcat}.

\smallskip\noindent\textbf{Dependence on the password-hashing scheme.}
The rank, and hence every entropy reduction reported in \S\ref{sec:userstudy}, is independent of how the password is stored: it is determined by the video evidence and the prior. Only the rank-to-time conversion involves the hashing scheme, through the cracking speed we denote by~$s$ [guesses/second]. A slow and memory-hard key-derivation function such as bcrypt, scrypt, or Argon2 reduces~$s$ by orders of magnitude relative to the fast-hash rate assumed here~\cite{huang2024probhashcat}. Since crack time is $r/s$, this rescales all our reported times by a constant factor, leaving the rank and the entropy reductions unchanged. We therefore report entropy bits alongside crack times throughout.

\section{The User Study} \label{sec:userstudy}
We evaluate our attack pipeline using real-world user studies under a diverse set of participants and conditions. We recruited 16~participants (P0--P15; mean age 37.1 years, $\text{SD} = 14.1$; 10 male, 6 female) to type various passwords (see \ref{para:password_test_set_para} below). 10 participants typed with two thumbs, one (P2) used a single thumb, and 5 participants 
typed with one index finger. 
Younger participants typed considerably faster and with fewer hesitations. The detailed participant information is in Table~\ref{tab:paricipants_data} in Appendix~\ref{app:participants_passwords}. 
All the experiments were conducted under our Institutional Review Board (IRB) approved protocol. We organize our results into five groups: (1)~performance across users, (2)~performance under varying scenarios, (3)~overall accuracy and failure analysis, (4)~ablation of pipeline components, and (5)~computational cost.

\subsection{Experimental Setup}\label{sec:setup}

\noindent\textbf{Password test set.}\label{para:password_test_set_para} 
To ensure consistency across participants, we compiled a fixed test set of 18~passwords. Nine are \emph{human-chosen} passwords of lengths 8--16, drawn randomly from the Jason corpus (905~million passwords) as provided in~\cite{david2021explainable}. The remaining nine are \emph{password-manager-chosen} random passwords of lengths 8--16, produced by Bitwarden's password generator~\cite{bitwarden}. All passwords satisfy our composition policy (minimum 8~characters, at least one uppercase letter, at least one digit or special character). The specific passwords used are in Table~\ref{tab:password_test_set} in Appendix~\ref{app:participants_passwords}.

\smallskip\noindent\textbf{Victim configuration.}
Participants were seated and asked to hold their smartphone naturally, ensuring only that the screen and fingers remain visible to the camera. Before filming, each participant practiced every password to reduce cognitive load, mimicking the fluency of typing a familiar password. The default device was a 6.3-inch Google Pixel~10 with Gboard, and we used an iPhone 16 and Galaxy A53 in specific scenarios. Table~\ref{tab:devices} in Appendix~\ref{app:devices} reports the operating-system and keyboard-application versions used for each smartphone. All keyboards were left in their default configuration, with no user personalization, and no changes to display-size or font-scaling settings.

\smallskip\noindent\textbf{Attacker configuration.}
The researcher playing the part of the attacker wore the Meta Ray-Ban smart glasses~\cite{meta-rayban} and sat approximately 0.8--1.0\,m from the victim, with the camera 40--60\,cm above the phone. The attacker tried to keep head movements minimal.

\smallskip\noindent\textbf{Metrics.}
\begin{itemize}
    \item \emph{Attack strength}: entropy bits, $\log_2(r)$, where $r$ is the password's estimated rank.
    \item \emph{Crack time}: $r / s$, where $s = 10^{10}$~guesses/second against an offline verifier storing a fast hash (\S\ref{sec:threat}). For example, a rank of $2^{45}$ yields a crack time of approximately one hour. Crack times scale linearly in~$s$. 
    \item \emph{Accuracy}: binary per-password - the attack succeeds if and only if the correct typed-sequence length appears among the candidate predictions. This is a \emph{keystroke-recovery} rate and not a rate of passwords cracked, as it measures how often the pipeline produces a candidate list that is usable at all. Attack-strength results are reported over the entries meeting this criterion, and \S\ref{sec:accuracy} relates the two.
\end{itemize}

\subsection{Performance Across Users}\label{sec:users}
We requested our 16 participants to type the 9 human-chosen passwords and the 9 password-manager-chosen passwords in our default attacker-victim configuration. In addition to the 18-password test set, each participant typed one additional 10-character human-chosen password (\texttt{1Piece4You}) three times: once without errors, once with one backspace correction, and once with two consecutive backspace corrections. This yielded 336~videos in total.

In what follows, we first present the attack strength results for all participants on whom the attack successfully detected all keystrokes. These results constitute the central contribution of this evaluation. We then analyze performance under varying physical scenarios, and finally discuss overall accuracy and failure modes.

\begin{figure}[t]
    \centering
    \begin{subfigure}{0.5\textwidth}
        \centering
        \includegraphics[width=\textwidth]{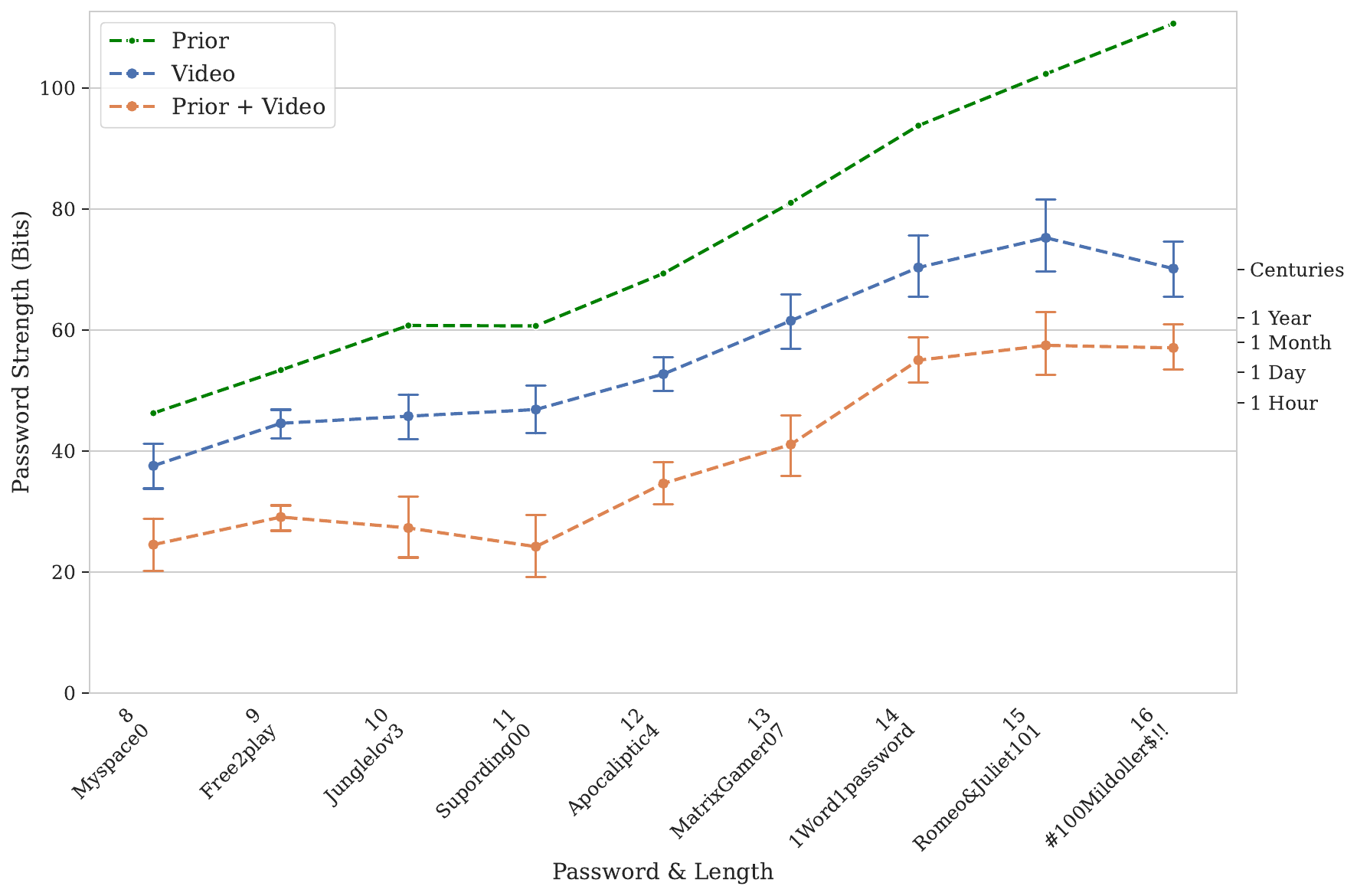}
        \caption{Human-chosen passwords attack strength.}
        \label{fig:all_password_common}
    \end{subfigure}
    \hfill
    \begin{subfigure}{0.5\textwidth}
        \centering
        \includegraphics[width=\textwidth]{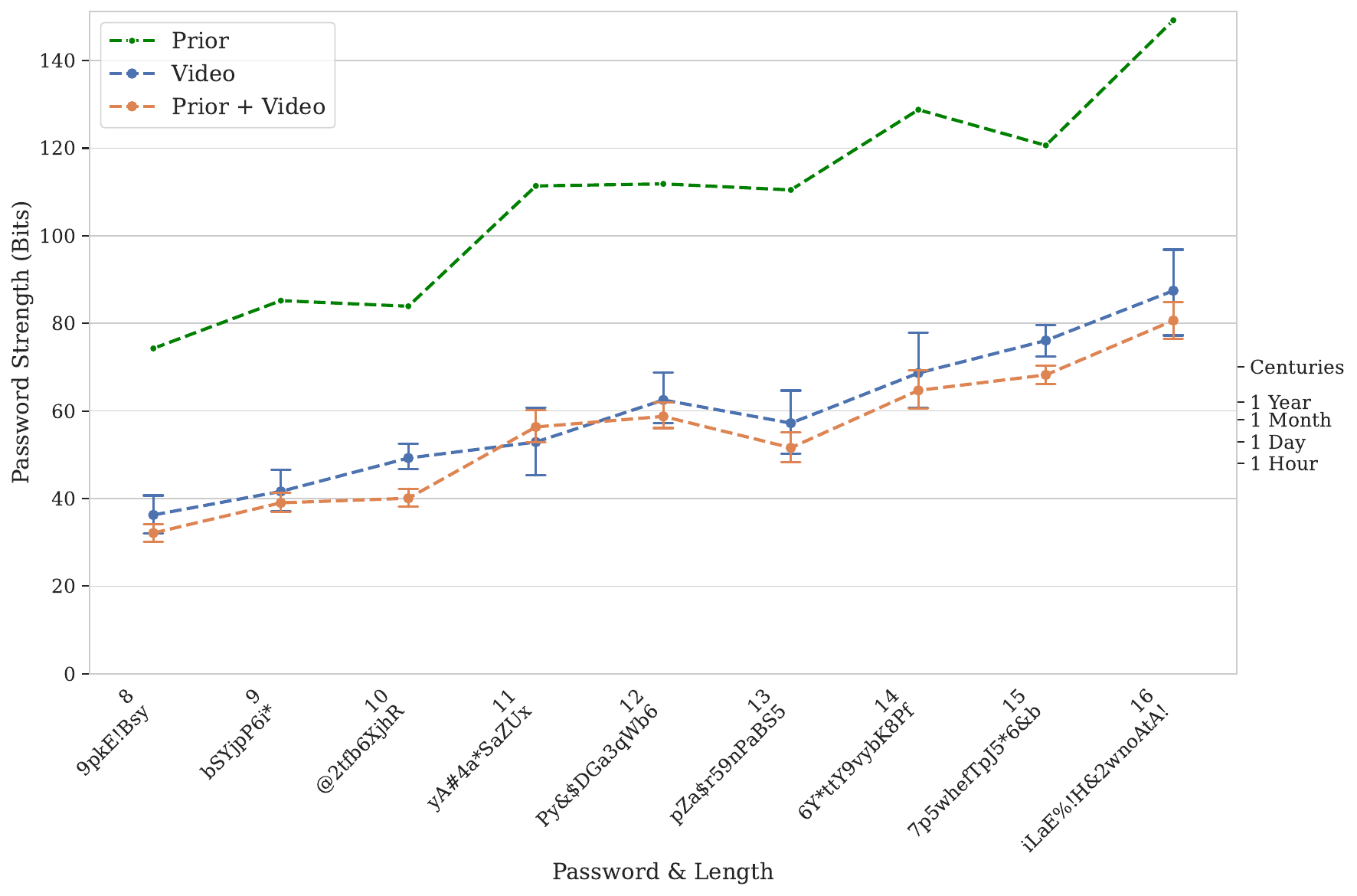}
        \caption{Password-manager-chosen passwords attack strength.}
        \label{fig:all_password_generated}
    \end{subfigure}
    \caption{Average attack strength for 18 passwords across all 16 participants, with 95\% confidence intervals, using the prior only, video only, and their conflated distribution.}
    \label{fig:all_passwords_results}
\end{figure}

\smallskip\noindent\textbf{Attack strength on human-chosen passwords.} Figure~\ref{fig:all_password_common} shows the prior, video, and conflated rank (in bits of entropy and approximated crack time) for the 9~human-chosen passwords across all users for whom the attack successfully detected all keystrokes. This is the central result of our evaluation and demonstrates the power of combining video evidence with prior password knowledge.

Three key observations emerge. First, the prior rank alone exceeds practical crack times (centuries) for human-chosen passwords of length~10 and above, confirming that prior knowledge alone is insufficient for long enough passwords. Second, the video data reduces the search space by 10--40~bits relative to the prior. Third, and most importantly, the \emph{conflated rank} of combining prior and video via the conflation operator (Equation~\ref{eq:conflation}) achieves the lowest rank across all passwords, reducing the search space by 20--60~bits relative to the prior alone. At a 95\% confidence interval, the video and conflated ranks are statistically distinguishable, with non-overlapping confidence intervals of approximately 20~bits across most password lengths. A paired Wilcoxon signed-rank test over per-user ranks confirms that the difference between the video-only and conflated distributions is statistically significant ($p < 0.001$) for all nine human-chosen passwords. This statistically significant gap demonstrates that an attacker who leverages both the video observation and publicly available password corpus data gains a substantial advantage over using either source alone. Under the conflated model, human-chosen passwords of up to 16~characters can be cracked in hours to days.

\smallskip\noindent\textbf{Attack strength on generated passwords.}
Figure~\ref{fig:all_password_generated} shows analogous results for password-manager-chosen passwords. Here the story is markedly different. The prior distribution is far less informative for generated passwords, as they do not follow human composition patterns: the prior rank exceeds 80~bits for passwords as short as 9 characters. Consequently, the conflated and video-only ranks largely coincide, with only a slight advantage for the conflated rank on average. This confirms that for password-manager-chosen passwords, the attacker must rely primarily on the video data. A paired Wilcoxon signed-rank test confirms that the difference between the video-only and conflated ranks is not statistically significant ($p > 0.05$) for most generated passwords, consistent with the visual overlap of confidence intervals in Figure~\ref{fig:all_password_generated}. Nevertheless, the video evidence alone is powerful: it reduces the search space by 40--70~bits relative to the prior. The figure shows that randomly generated passwords of up to 13~characters become crackable in under an hour. From 14~characters onward, crack times extend to centuries even with video data.

Not surprisingly, the contrast between these two results quantifies the common insight that human-chosen passwords, even when satisfying composition rules, are significantly more vulnerable to attack than password-manager-chosen passwords of the same length, because the prior distribution provides the attacker with an advantage of 10--20~bits.

\begin{figure}[t]
    \centering
    \includegraphics[width=1.0\columnwidth]{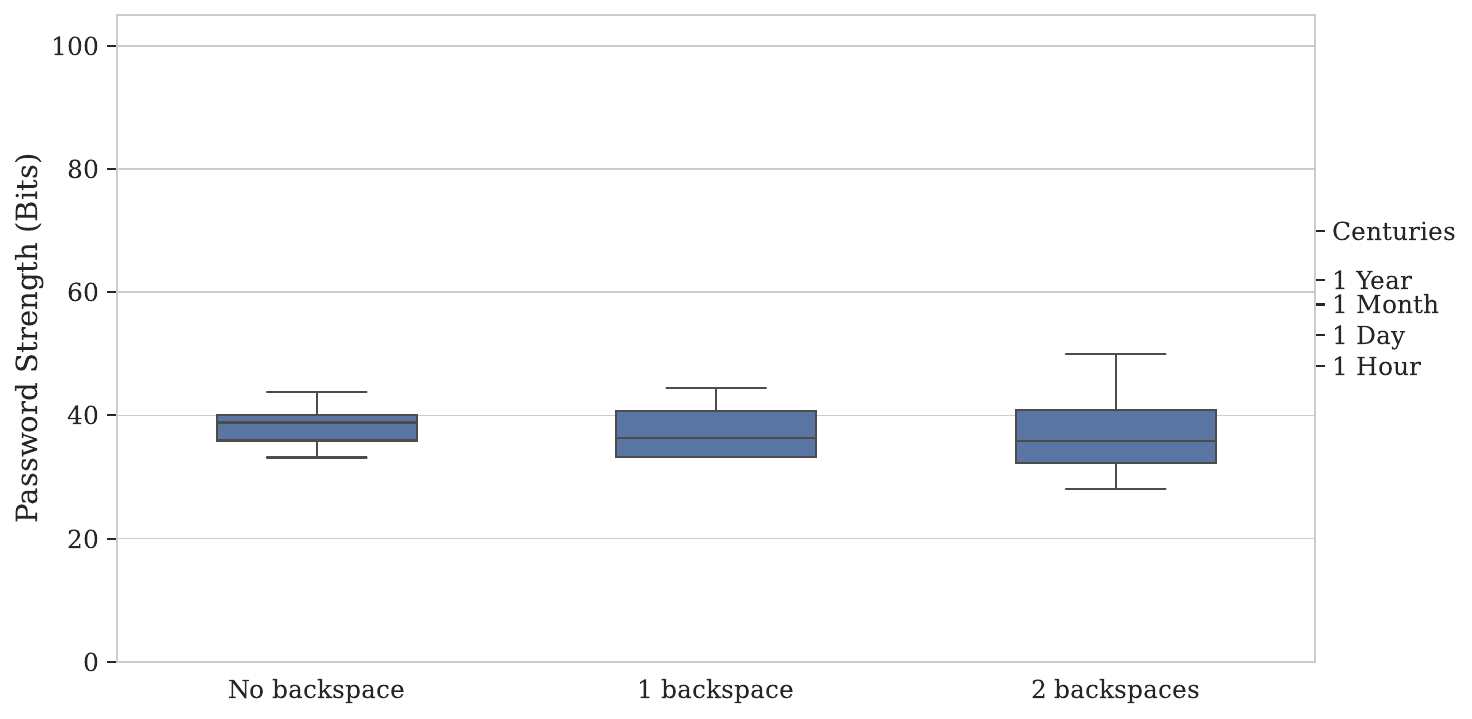}
    \caption{Attack strength when deleting characters with the backspace key for the password (\texttt{1Piece4You}).}
    \label{fig:backspace}
\end{figure}

\smallskip\noindent\textbf{Backspace robustness.}
Figure~\ref{fig:backspace} shows the conflated rank when participants type a 10-character password with zero, one, or two backspace corrections. The rank remains comparable across all three conditions, confirming that successful backspace detection preserves the effective typed-sequence length and does not degrade attack performance.

\subsection{Performance Under Varying Scenarios}\label{sec:scenarios}

Beyond the default setup, we evaluated the attack's performance in additional configurations using two participants, one thumb typist and one index-finger typist. We considered two configuration types: the physical capture configuration of the attacker--victim pair, and the smartphone device used by the victim.

We first evaluated physical configurations by fixing a single 10-character password-manager-chosen password (\texttt{@2tfb6XjhR}, typed-sequence length~16) and varying three dimensions: distance (4~levels), camera elevation (5~levels), and viewing angle (3~levels). Each participant repeated the password 8~times per configuration, yielding 16~videos per configuration. Because the baseline configuration (0.8\,m, 50\,cm, frontal view) is shared across all three dimensions, the number of unique physical configurations is $(4+5+3)-2=10$, giving $10 \times 16 = 160$ physical configuration videos.

We then evaluated device variability by recording the entire 18-password test set on two additional smartphones beyond the Google Pixel~10 already used in \S\ref{sec:users}, with each participant typing each password once. This adds $2\times 18\times 2 = 72$ device-comparison videos. The combined scenario corpus thus comprises $160+72=232$ unique videos. All results below report conflated rank.

\begin{figure}[t]
    \centering
    \includegraphics[width=1.0\columnwidth]{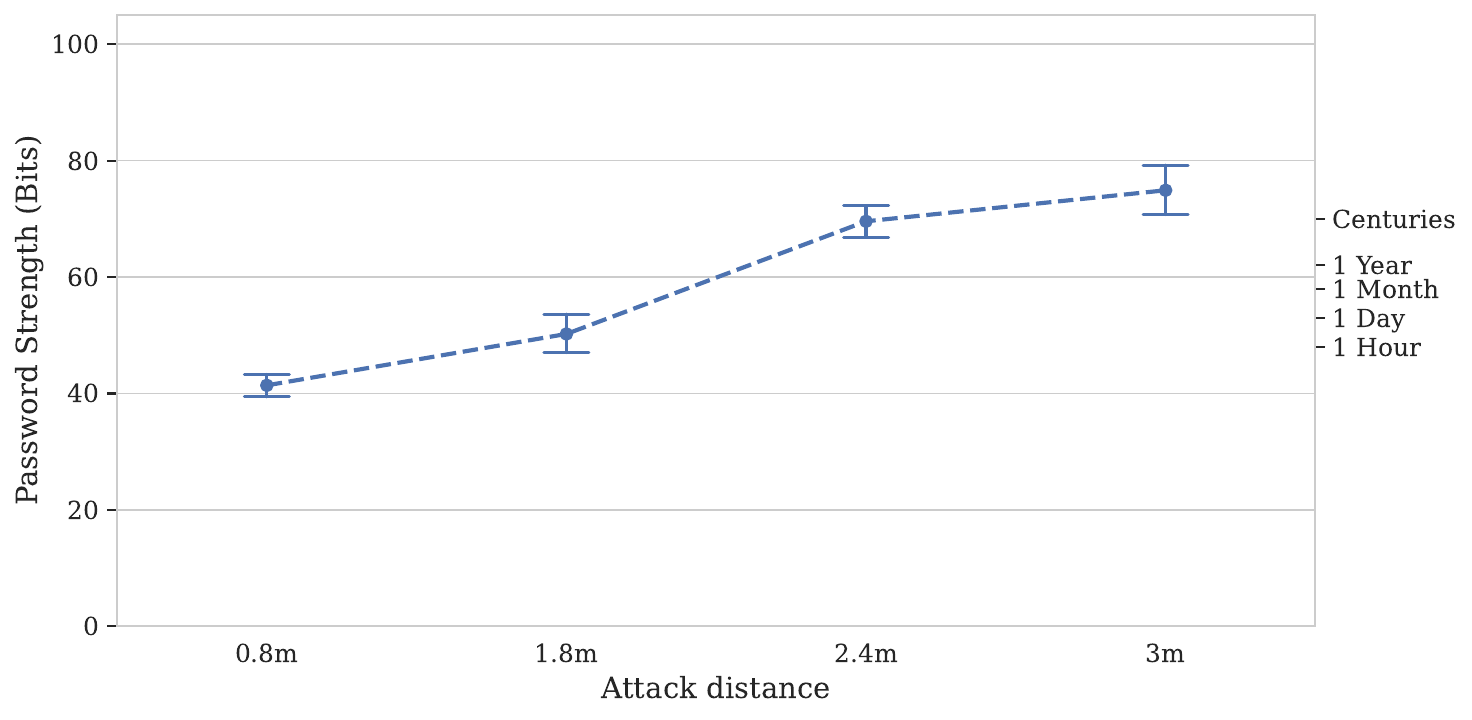}
    \caption{Attack strength when varying the distance.}
    \label{fig:distance}
\end{figure}

\smallskip\noindent\textbf{Scenario 1: Distance.}
We varied the attacker victim distance from 0.8\,m to 3.0\,m (camera height fixed at~50\,cm above the phone). As shown in Figure~\ref{fig:distance}, the attack remains effective up to 1.8\,m. Beyond this, subtle head movements of the attacker increasingly degrade the captured footage. We note that the default Meta Ray-Ban camera has no zoom capability and optical zoom would likely extend the effective range.

\begin{figure}[t]
    \centering
    \includegraphics[width=1.0\columnwidth]{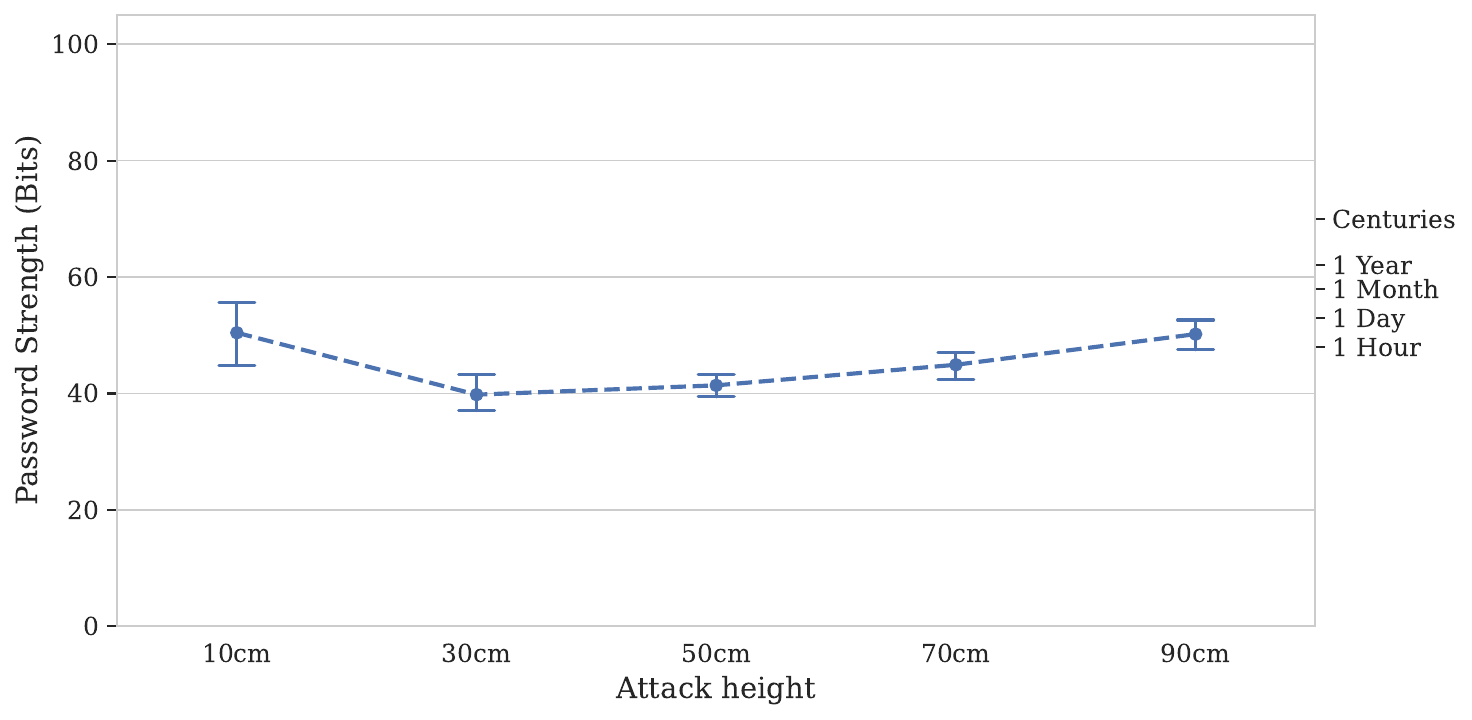}
    \caption{Attack strength when varying the height.}
    \label{fig:height}
\end{figure}

\smallskip\noindent\textbf{Scenario 2: Camera elevation.}
We varied the vertical offset between the camera and the phone from 10\,cm to 90\,cm (distance fixed at 0.8\,m). These elevations correspond to common attacker--victim posture combinations: both seated, both standing, or the attacker standing while the victim is seated. We simulated different elevations by having the victim hold the smartphone at varying positions on their upper body (e.g., chest height or near the waist). Figure~\ref{fig:height} shows that the attack is effective across all tested elevations, with a sweet spot at 30--50\,cm, corresponding to sitting--sitting or standing--standing configurations, where the victim holds the device anywhere between the chest and waist. At very low elevations ($\leq$10\,cm), the nearly flat viewing angle makes screen-corner estimation difficult. At high elevations ($\geq$70\,cm), the finger's contact point with the screen becomes harder to resolve, despite the screen itself being more clearly visible.

\begin{figure}[t]
    \centering
    \includegraphics[width=1.0\columnwidth]{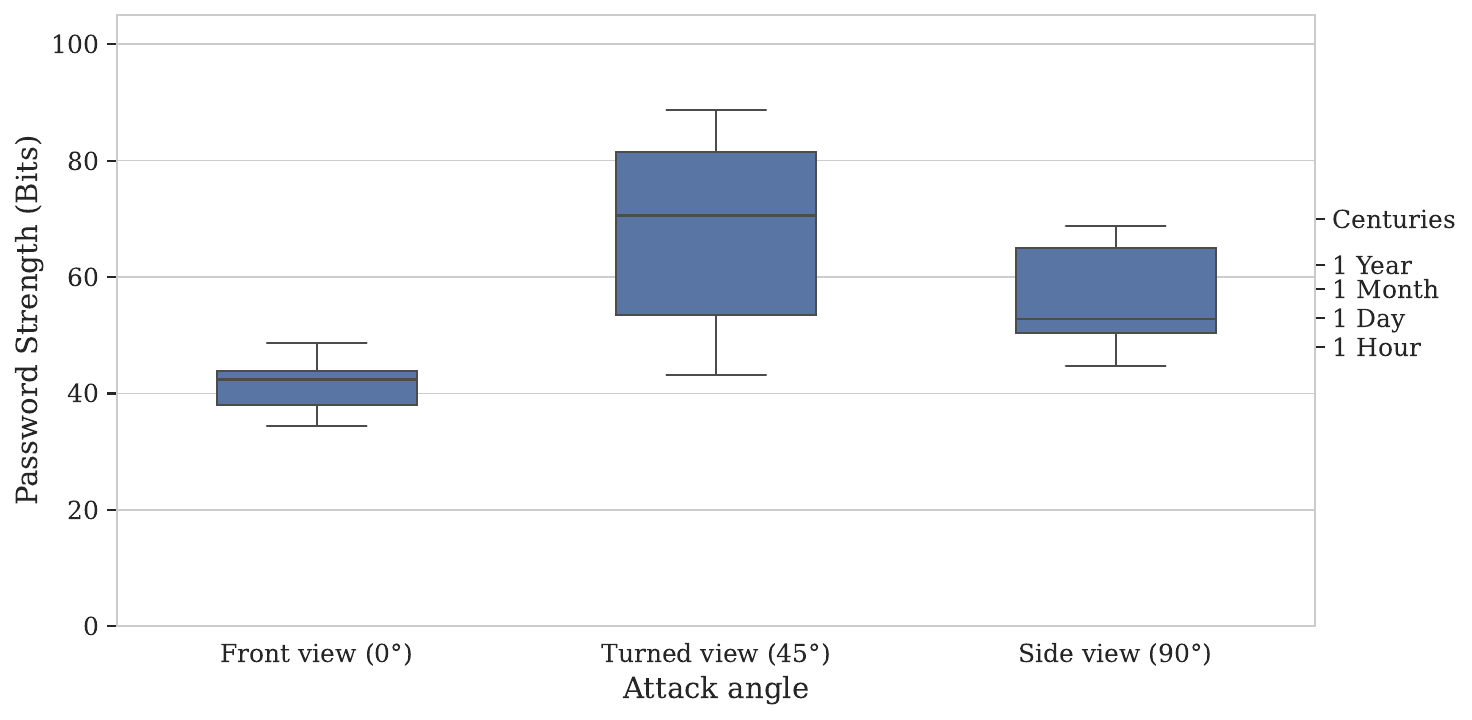}
    \caption{Attack strength when varying the angle.}
    \label{fig:angle}
\end{figure}

\smallskip\noindent\textbf{Scenario 3: Viewing angle.}
We tested frontal (0\textdegree), oblique (45\textdegree), and side-profile (90\textdegree) views (distance 0.8\,m, elevation 50\,cm). Figure~\ref{fig:angle} reports results for different views the attacker has to the screen. At 0\textdegree, results are consistent and the password is crackable. Both 45\textdegree\ and 90\textdegree\ views produced degraded results. At 45\textdegree, we attribute this to jitter in SAM\,3's screen segmentation at oblique angles. At 90\textdegree, the screen segmentation is more stable at a side view, but performance sometimes collapsed due to one finger occluding the other and inconsistent tracking. We conclude that while the attack is possible at varying viewing angles, it is most effective at a front view which enables clear keyboard estimation and finger tracking.

\begin{figure}[t]
    \centering
    \begin{subfigure}[b]{0.5\textwidth}
        \centering
        \includegraphics[width=\textwidth]{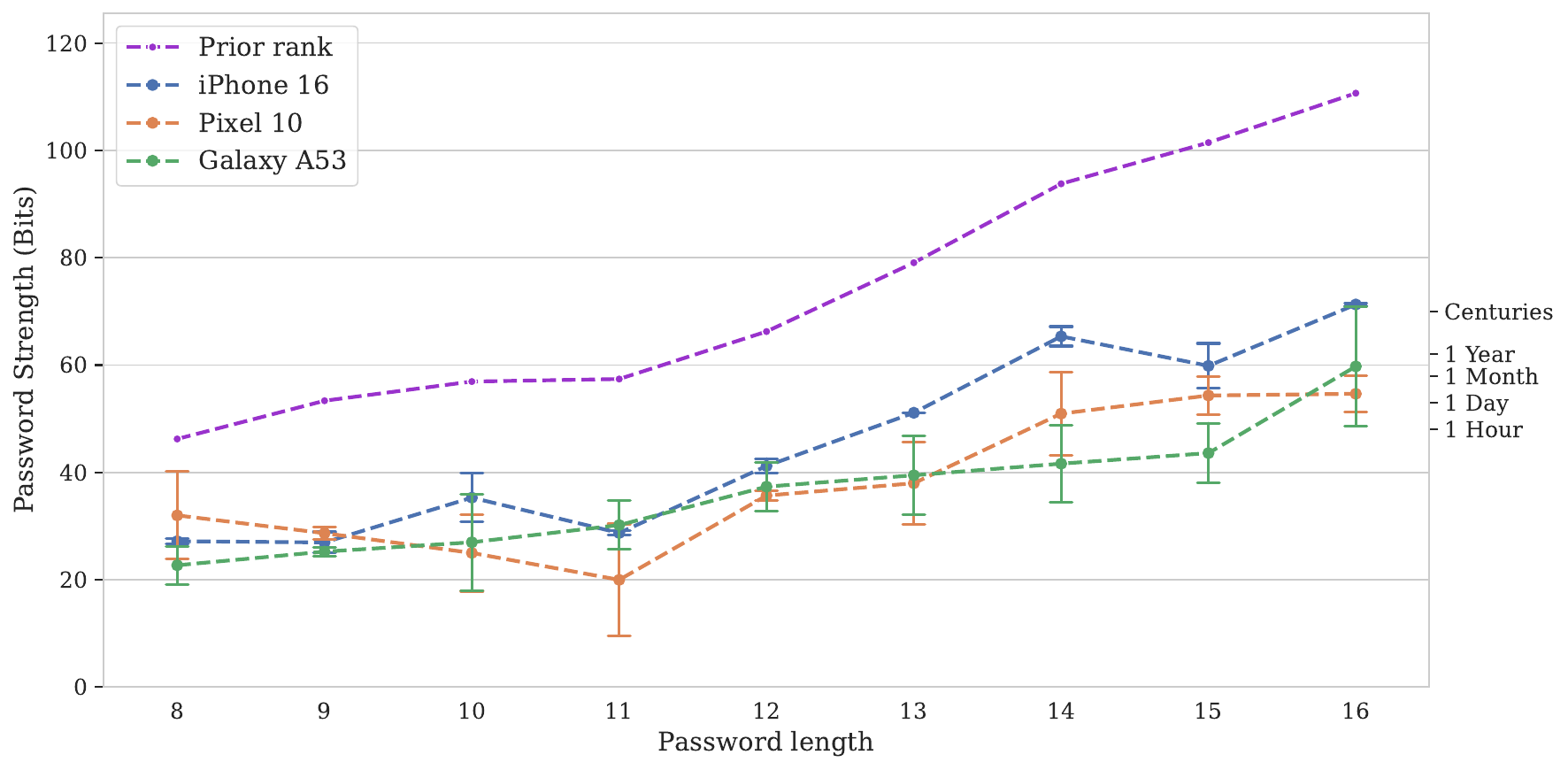}
        \caption{Human-chosen passwords attack strength.}
        \label{fig:device-common}
    \end{subfigure}
    
    \vspace{0.2cm}
    
    \begin{subfigure}[b]{0.5\textwidth}
        \centering
        \includegraphics[width=\textwidth]{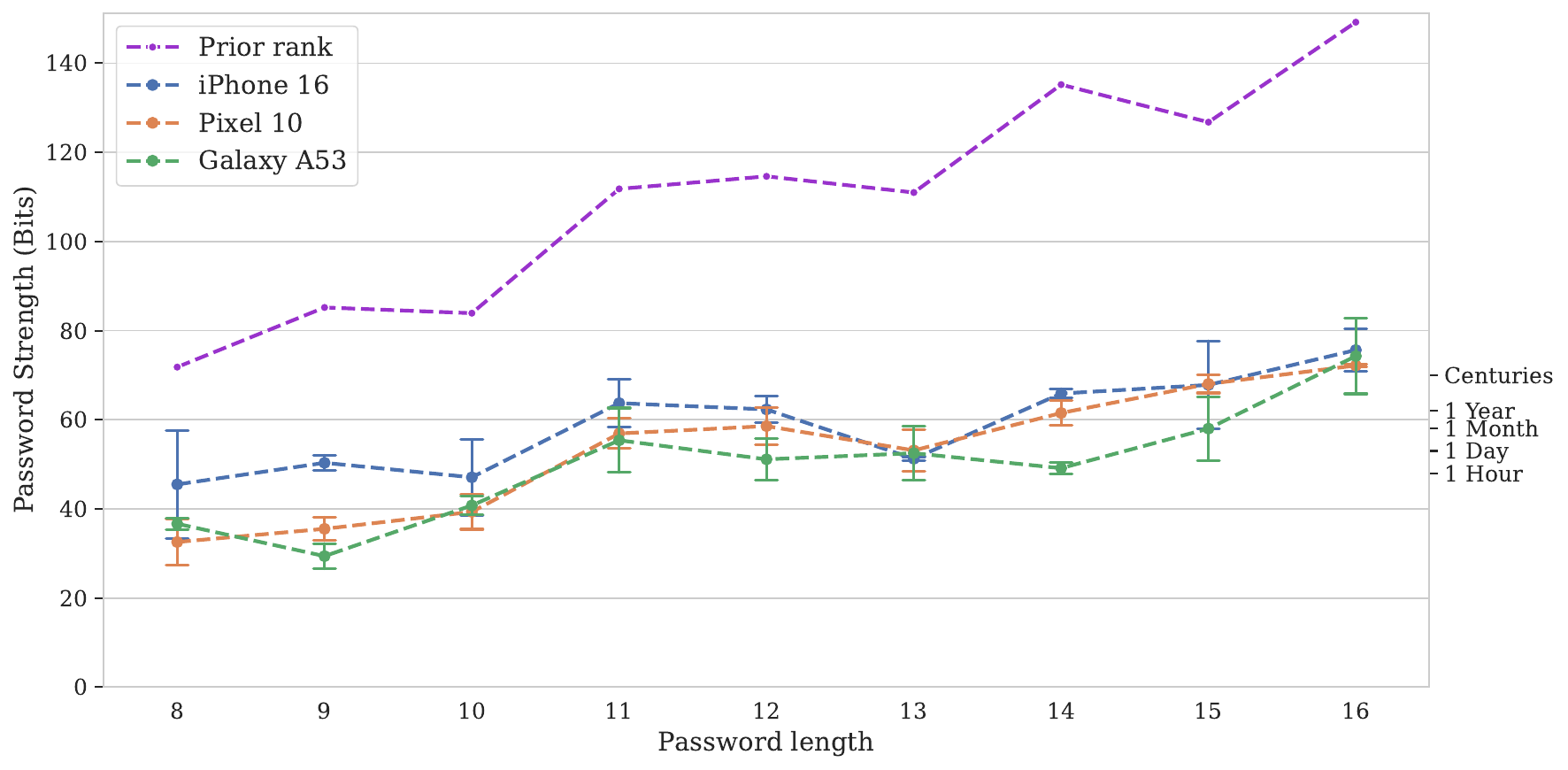}
        \caption{password-manager-chosen random passwords attack strength.}
        \label{fig:device-generated}
    \end{subfigure}
    
    \caption{Attack strength when typing password test set on iPhone 16, Pixel 10, and Samsung Galaxy A53.}
    \label{fig:all_phones_results}
\end{figure}

\smallskip\noindent\textbf{Scenario 4: Device variability.}
We repeated the full 18-password test set on three devices: Google Pixel~10 (6.3\,in, Gboard), iPhone~16 (6.1\,in, iOS keyboard), and Samsung Galaxy~A53 (6.5\,in, Samsung keyboard). Figures~\ref{fig:device-common} and~\ref{fig:device-generated} show conflated ranks across devices. The attack generalizes across all three, validating the assumption-free design. A paired Wilcoxon signed-rank test over all pairwise device combinations confirms that the differences in conflated rank across devices are not statistically significant. Nevertheless, consistent patterns emerge across devices. As noted in \S\ref{sec:typing}, the Samsung keyboard's five-row layout places digits directly above the letter rows, reducing the required touch count for passwords containing numbers. In our multi-configuration pipeline, the Samsung template therefore produces shorter typed sequences for many passwords, which removes some uncertainty and yields slightly lower ranks. Conversely, the iOS keyboard places several common symbols in a deeper layout layer than Gboard does, requiring additional layout switches and increasing the rank. We observe that keyboards minimizing layout transitions inadvertently reduce password security against this attack.

\subsection{Overall Accuracy and Failure Analysis}\label{sec:accuracy}

\begin{figure}[t]
    \centering
    \includegraphics[width=1.0\columnwidth]{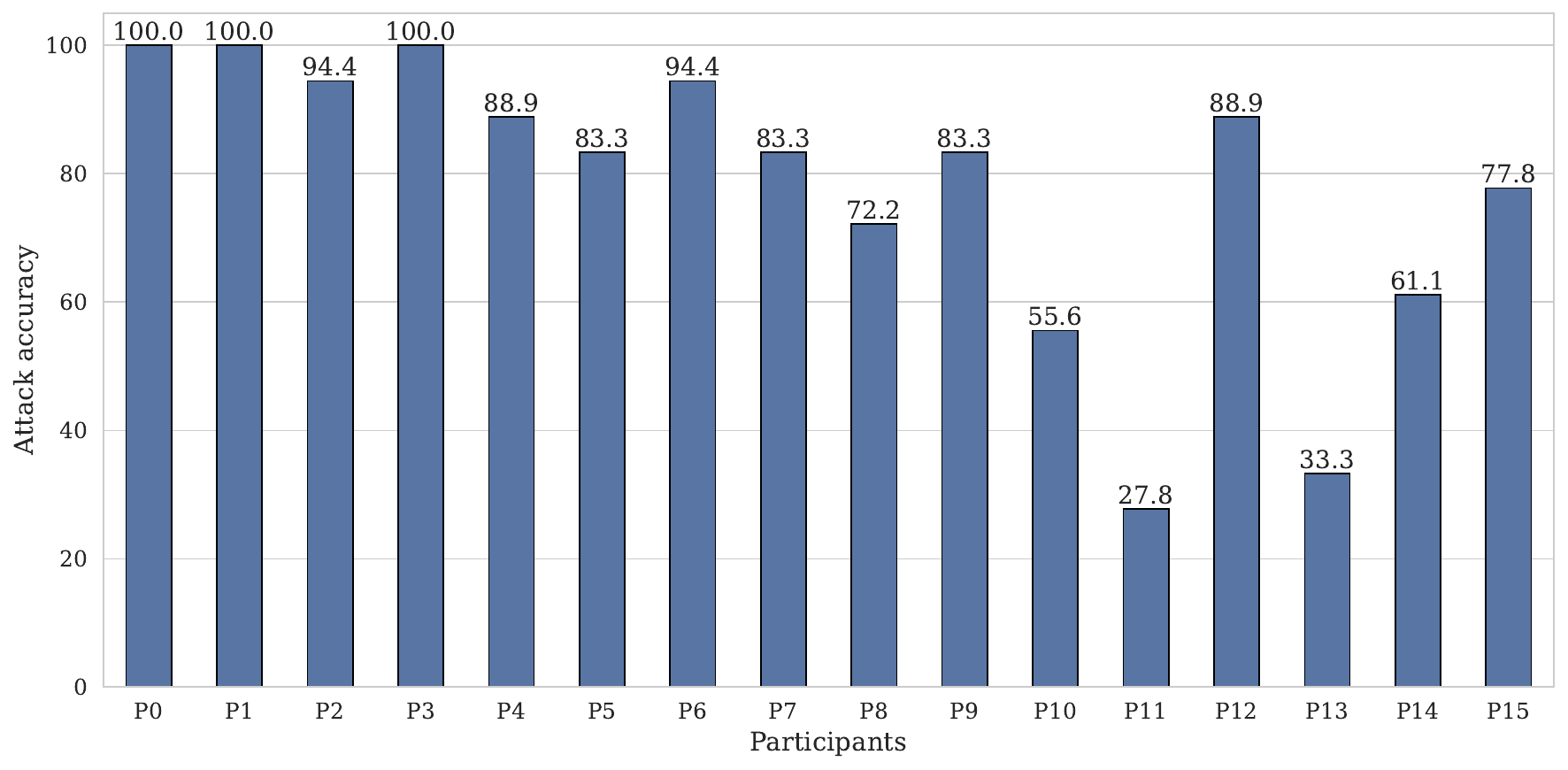}
    \caption{Attack accuracy for all 16 participants.}
    \label{fig:attack_users_accuracy}
\end{figure}

The attack has two stages whose success criteria must be kept distinct. The pipeline must first recover the exact typed-key sequence, and only then does the rank reported in \S\ref{sec:users} describe a candidate list that the adversary can actually enumerate. If a keystroke is missed or spuriously inserted, the recovered sequence has the wrong length and that attempt's rank is not a usable cracking effort. The attack strength results above are therefore conditioned on the pipeline successfully detecting all keystrokes, a strict all-or-nothing criterion under which a single missed or spurious keystroke constitutes a failure. Figure~\ref{fig:attack_users_accuracy} reports the per-user success rate across the full 18-password test set. The overall accuracy is \textbf{77.8\%}, which is the rate of clearing this first stage rather than the fraction of passwords cracked, and is not a discount to be applied to the crack times of \S\ref{sec:users}. For three users (P0, P1, P3), the attack achieved 100\% accuracy, while four others (P10, P11, P13, P14) had the attack fall below 65\%.

Analysis of the failure cases reveals distinct causes tied to typing behavior:

\begin{itemize}
    \item \textbf{Fast, subtle typists (P11, P13).} Against these two participants, the attack achieved only 27.8\% and 33.3\% accuracy respectively. They are the fastest typists in our study, with very subtle finger movements (1--2\,cm travel between keystrokes). The GMM-based touch estimation failed to detect peaks with sufficient prominence, resulting in too few high-confidence pseudo-labels to effectively train the CNN ensemble. Notably, these are the youngest participants, suggesting that attack efficacy may degrade for younger, faster typists.
    
    \item \textbf{Near-simultaneous touches (P10).} At 55.6\% accuracy, P10 is also a younger, fast typist who frequently exhibited near-simultaneous keypresses with both thumbs, making it difficult for the pipeline to distinguish individual keystrokes.
    
    \item \textbf{Hesitant typing (P14).} Unlike the above cases, P14 is an older, single-finger typist who typed slowly but exhibited frequent hesitations between presses. These hesitations produced high-prominence false peaks that confused the CNN ensemble, generating false positives.
    
    \item \textbf{Effect of password complexity.} We also observe that success rates decline with password length and complexity (graph omitted). The 8-character human-chosen password was successfully predicted for 93.8\% of users, while the 16-character random password was successfully predicted for only 50.0\%. This is expected: longer passwords introduce more keystrokes and thus more opportunities for a single detection error to invalidate the entire prediction.
    
\end{itemize}

\subsection{Component Ablation}\label{sec:ablation}

Table~\ref{tab:ablation} reports the effect of removing key pipeline components for two representative participants: P3 (two-thumb typist, 100\% overall accuracy) and P8 (index-finger typist, 72.2\% overall accuracy).

\noindent\textbf{SAM\,3 finger tracking.}
Without SAM\,3, P3 drops to 5.5\% because MediaPipe fails almost entirely on one hand (palm occlusion). SAM\,3 recovers by segmenting fingers directly via concept prompts. For P8, removing SAM\,3 reduces accuracy by approximately 28 percentage points.

\smallskip\noindent\textbf{CNN ensemble.}
Without the ensemble, i.e., relying solely on GMM-based touch estimation, accuracy drops to 11.1\% (P3) and 27.8\% (P8). The GMM frequently misses low-prominence keystrokes that the CNN, trained on the high-confidence subset, can recover.

\begin{table}[t]
\centering
\small
\setlength{\tabcolsep}{1.5pt}
\begin{tabular}{@{}cccc|cc@{}}
\toprule
MediaPipe & GMM touch & SAM\,3 fingers & CNN ensemble & P3 Acc. & P8 Acc. \\
\midrule
\cmark & \cmark & & \cmark & 5.5\% & 44.4\% \\
\cmark & \cmark & \cmark & & 11.1\% & 27.8\% \\
\cmark & \cmark & \cmark & \cmark & 100\% & 72.2\% \\
\bottomrule
\end{tabular}
\caption{Ablation study: contribution of SAM\,3 finger tracking and CNN ensemble to attack accuracy.}
\label{tab:ablation}
\end{table}

\begin{figure}[t]
    \centering
    \includegraphics[width=1.0\columnwidth]{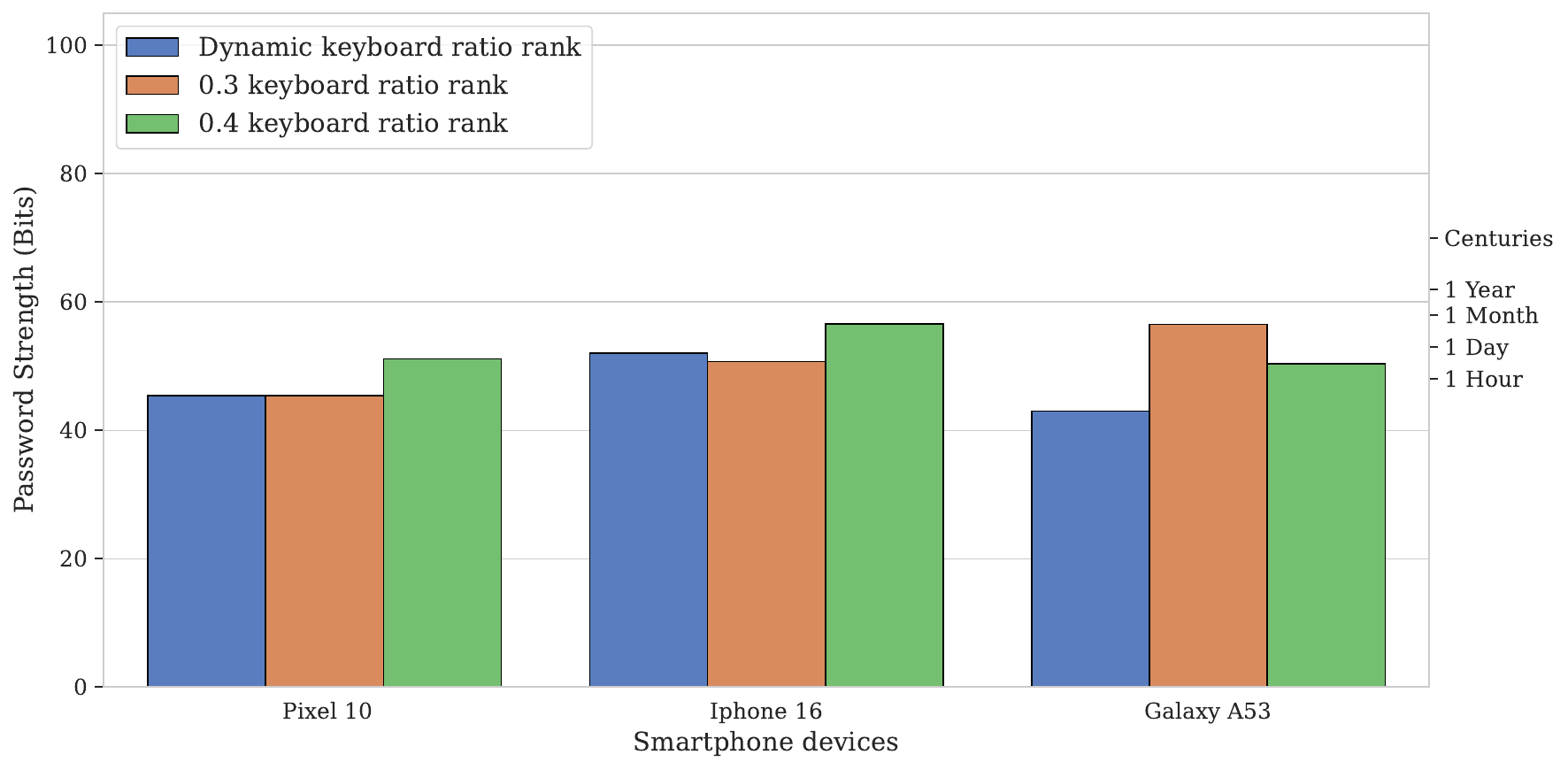}
    \caption{Attack  strength for different keyboard ratios.}
    \label{fig:ratio}
\end{figure}

\smallskip\noindent\textbf{Dynamic keyboard ratio.}
We also compared our dynamic keyboard estimation against two fixed ratios, $0.3$ and $0.4$, that correspond to typical fractions of the screen height occupied by the on-screen keyboard on popular smartphones. We averaged the conflated rank across the entire 18-password test set for each smartphone device. In Figure~\ref{fig:ratio}, a lower bar indicates a lower estimated rank, and therefore a stronger attack. The fixed ratios yield inconsistent performance across devices (0.3 favors iPhone; 0.4 favors Samsung), whereas the dynamic approach achieves consistently competitive ranks across all three smartphones without requiring prior knowledge of the keyboard geometry.

\begin{figure}[t]
    \centering
    \includegraphics[width=1.0\columnwidth]{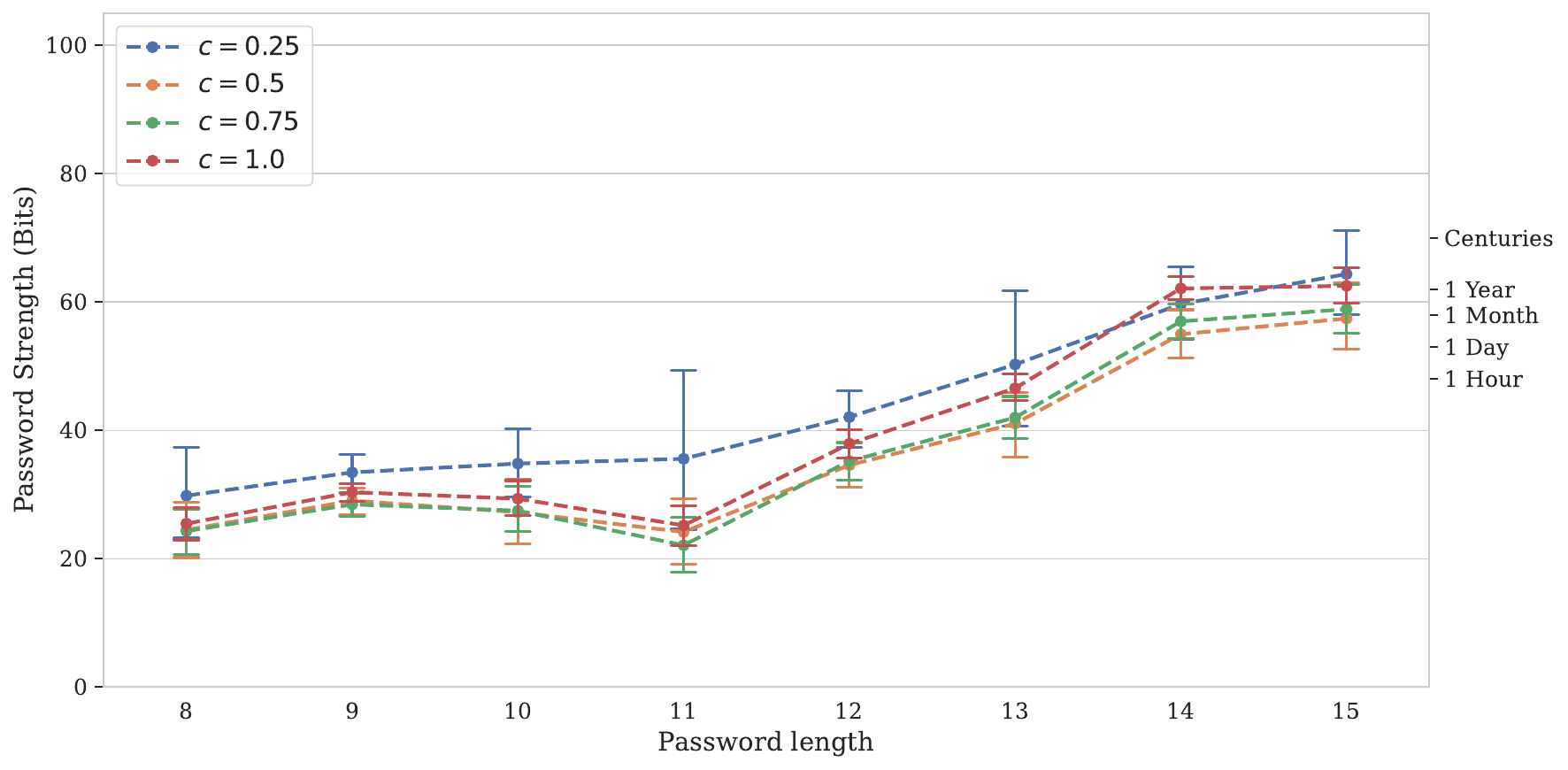}
    \caption{Attack strength for different $\sigma$ values.}
    \label{fig:sigma_ablation}
\end{figure}

\smallskip\noindent\textbf{Gaussian sigma sensitivity.}\label{sec:sigma_choice}
Finally, we evaluated the sensitivity of the attack to the standard deviation coefficient used in the per-key probability model (recall \S\ref{sec:step6}). We tested four values of the coefficient $c \in \{0.25, 0.5, 0.75, 1.0\}$, where $\sigma_x = c \cdot w_{\min}$ and $\sigma_y = c \cdot h$. Figure~\ref{fig:sigma_ablation} shows the conflated rank across all nine human-chosen passwords for all 16 participants. The results are robust for $c \in \{0.5, 0.75, 1.0\}$, with rank differences of only a few bits among these values. Performance degrades sharply at $c = 0.25$, where the Gaussian is too narrow and assigns near-zero probability to the correct key when the touch estimate is slightly off-center. The optimal coefficient lies near $c = 0.5$, consistent with the geometric intuition that a touch at a key boundary should fall approximately one standard deviation from the key center.

\subsection{Computational Cost}\label{sec:cost}

We benchmark the pipeline on a PC with an Intel Core i9-12900KF CPU and NVIDIA RTX~3080~Ti GPU. For a 1-minute attack video (a single password of length 13+), the total processing time is 10--12~minutes: phone segmentation ($\sim$1\,min), MediaPipe tracking ($\sim$5\,s at 30\,FPS), SAM\,3 finger segmentation and tracking ($\sim$5\,min), touch estimation ($\sim$5\,s), CNN ensemble training ($\sim$5\,min for five networks), and keyboard estimation, probability computation, and rank estimation ($<$10\,s).

\section{Discussion} \label{sec:discussion}

Before discussing defenses and limitations, we briefly contextualize the scale of our evaluation relative to prior work. Our participant count (16) matches that of Yang~et~al.~\cite{yang2023general}, the most closely related general video-based keystroke inference attack, while targeting a substantially harder problem: full-QWERTY password inference on smartphones rather than text inference on tablets. Our experimental distances (0.8--3.0\,m) also mirror theirs, enabling direct methodological comparison. Our total corpus of 568~unique videos (336~from the cross-user study plus 232~from the scenario experiments) compares favorably to prior video-based password attacks: Shukla and Phoha~\cite{shukla2019stealing} collected 375~videos across 45~volunteers but used only 3~videos per volunteer for testing and restricted scenario evaluation to 5~users with 42~videos each.

\subsection{Defenses}\label{sec:defenses}

Our results demonstrate that an opportunistic attacker using only smart glasses with a standard camera can crack passwords in realistic settings. We discuss several countermeasures:

\smallskip\noindent\textbf{Blocking line of sight.}
The most effective defense is to physically obstruct the attacker's view of the typing fingers. Tilting the smartphone toward the body so that the screen and fingers face upward rather than outward would render this attack ineffective. Conversely, privacy screen protectors alone do \emph{not} mitigate the attack as it relies on finger movements rather than screen content. A privacy display mode that activates on credential entry is weaker still, since its activation can itself signal when to record (\S\ref{sec:threat}).

\smallskip\noindent\textbf{Introducing ambiguity via dummy key presses.}
A user can deliberately press the layout-switch key multiple times before typing the actual password characters. Because the attack cannot distinguish meaningful layout switches from redundant ones, each extra press increases the number of candidate typed sequences an attacker must consider, multiplicatively increasing the rank. Unlike accidental hesitations, these are real touches to real keys that do not alter the entered password.

\smallskip\noindent\textbf{Longer and more random passwords.}
Our results confirm the intuitive expectation that longer and more random passwords are harder to crack. Generated passwords of 14+ characters remain secure even under a successful attack.

\smallskip\noindent\textbf{Alternative authentication.}
Biometric authentication (fingerprint, face recognition) and passkeys entirely eliminate the typing side channel for the authentication step---however they do not protect other sensitive text entry.

\smallskip\noindent\textbf{Two-factor authentication and attempt limiting.}
Two-factor authentication is a strong and recommended account-level mitigation which prevents account takeover even when the password is recovered in full. However, it does not neutralize our attack: it protects the account rather than the password, so a cracked password is still useful to the attacker if it is reused on a service lacking a second factor. 
Most importantly, two-factor authentication and attempt limiting are both online controls, which restrict login attempts against a live service but have no effect on offline guessing against a leaked hash (\S\ref{sec:threat}), which is the regime our crack times describe.

\subsection{Limitations and Future Work}\label{sec:limitations}

\noindent\textbf{Dependence on segmentation and tracking accuracy.}
Our pipeline relies on the current state of hand-tracking (MediaPipe) and segmentation (SAM\,3) models. Fast typists with minimal finger travel can cause the GMM and CNN stages to fail. Future improvements in real-time hand tracking and concept-aware segmentation may improve attack efficacy.

\smallskip\noindent\textbf{Public-space and motion conditions.}
Our evaluation is a controlled feasibility study rather than an in-the-wild demonstration, and it assumes a relatively stationary attacker and victim. We have not characterized crowd movement, outdoor or low-light illumination, or occlusion by sleeves, bags, or third parties. In active settings, such as a moving train or a walking attacker or victim, head and body movement introduces additional noise that our touch-estimation heuristics do not model. All of these degrade precisely the quantities to which the pipeline is most sensitive, namely screen-corner stability and fingertip localization, so we would expect accuracy to fall accordingly. The experiments of \S\ref{sec:scenarios} probe those sensitivities but do not substitute for a field study.

\smallskip\noindent\textbf{Phone posture and natural entry behavior.}
A victim who tilts the smartphone toward the body, types below table height, or shields the screen with the free hand removes the view the attack depends on. We did not systematically evaluate such postures. Furthermore, participants typed practiced, research-selected passwords while seated. This controls for memorization across participants but differs from entry of a long-memorized personal credential, and excludes autofill, password-manager copy-paste, and biometric unlock, each of which bypasses manual entry entirely and defeats the attack. Developing inference that is based on dorsal hand movement alone, without finger-to-screen visibility, remains an open challenge.

\smallskip\noindent\textbf{Non-QWERTY keyboards.}
We assume the standard QWERTY layout. Swipe-based input, alternative layouts (e.g., Dvorak), or custom keyboards are not currently handled. Long-press entry of digits and symbols is likewise unhandled.

\smallskip\noindent\textbf{Beyond passwords.}
Our pipeline targets short, policy-constrained strings. Extending it to general sensitive text entry would relax that length constraint while also admitting language-model correction that is unavailable for passwords~\cite{yang2023general}. Combining the two regimes would require deciding which one applies to a given session, which we leave to future work.

\section{Conclusions} \label{sec:conclusion}
We have presented the first general, video-based keystroke inference attack capable of estimating the rank of rule-based passwords adhering to realistic password policies typed on smartphones, without assumptions about the victim or their device beyond a frontal view of the typing hand and screen. Our pipeline leverages concept-aware segmentation for sub-pixel finger tracking, self-supervised 3D~CNN ensembles for keystroke prediction, dynamic keyboard estimation, and the ESrank algorithm for rank computation. Through user studies, we demonstrated that human-chosen passwords of up to 16~characters can be cracked in hours to days, and password-manager-chosen random passwords of up to 13~characters in under an hour, reducing up to 60~bits in password entropy compared to the prior baseline. A central finding is that conflating video-derived probabilities with prior password statistics yields a statistically significant reduction of approximately 20~bits in password entropy for human-chosen passwords, compared to either source alone. Our scenario analysis confirms robustness across distances up to 1.8\,m, a range of camera elevations and viewing angles, and three popular smartphone models. These findings underscore the practical privacy risk of typing passwords in public spaces in an era of ubiquitous wearable cameras, and motivate increased user awareness and defensive practices.

\bibliographystyle{ACM-Reference-Format}
\bibliography{references}

\appendix

\section{Ethical Considerations} 
This paper includes experimentation with human subjects: our test subjects were recorded on video as they typed passwords, which could pose risks to their privacy.  To mitigate the risk to the subjects we took the following steps:
\begin{itemize}
    \item The subjects all gave their consent to participating in the study.
    \item All video data was de-identified. Participants were assigned unique numeric IDs, and no personally identifiable information (PII) was linked to the raw video files.
    \item Audio recording on the smart glasses was muted, as audio is not required for the research.
    \item Recording data was stored on a secure local drive accessible only to the research team.
    \item The research team provided the phones on which the subjects typed to minimize the chance of leaking any personal data.
    \item The subjects were instructed to type passwords selected by the research team to eliminate the chance of their personal passwords being recorded.
    \item Screenshots of the attack overview included within this paper show one of the researchers and a separate individual who did not participate in the study, and did not type sensitive content during filming. Both signed approval to be published as part of the experiment setup.
\end{itemize}
We submitted our research protocol to our institutional Ethics Review Board (IRB) prior to conducting the experiments and received the IRB approval.

We recognize that this work demonstrates a practical attack that could be misused. We disclose it to raise awareness among users, keyboard developers, and the security community, and to motivate the development of defensive measures. The attack pipeline relies entirely on publicly available tools and models (MediaPipe, SAM\,3, R3D-18). We do not release the integrated attack code publicly, in order to prevent turnkey misuse. We have not conducted this attack against any non-consenting individuals. All recordings were made in controlled settings with full participant awareness.

Beyond IRB compliance, we considered the dual-use nature of this research. We believe the benefits of informing users of a concrete, realistic threat and providing actionable defensive guidance outweigh the risks, particularly given that the underlying tools are already publicly available and the attack scenario (recording someone in public) is already feasible without our specific pipeline.

\section{Smartphone Layouts}\label{app:layouts}
\begin{figure}[htbp]
    \centering
    
    \begin{subfigure}[b]{0.45\linewidth}
        \centering
        \includegraphics[width=\textwidth]{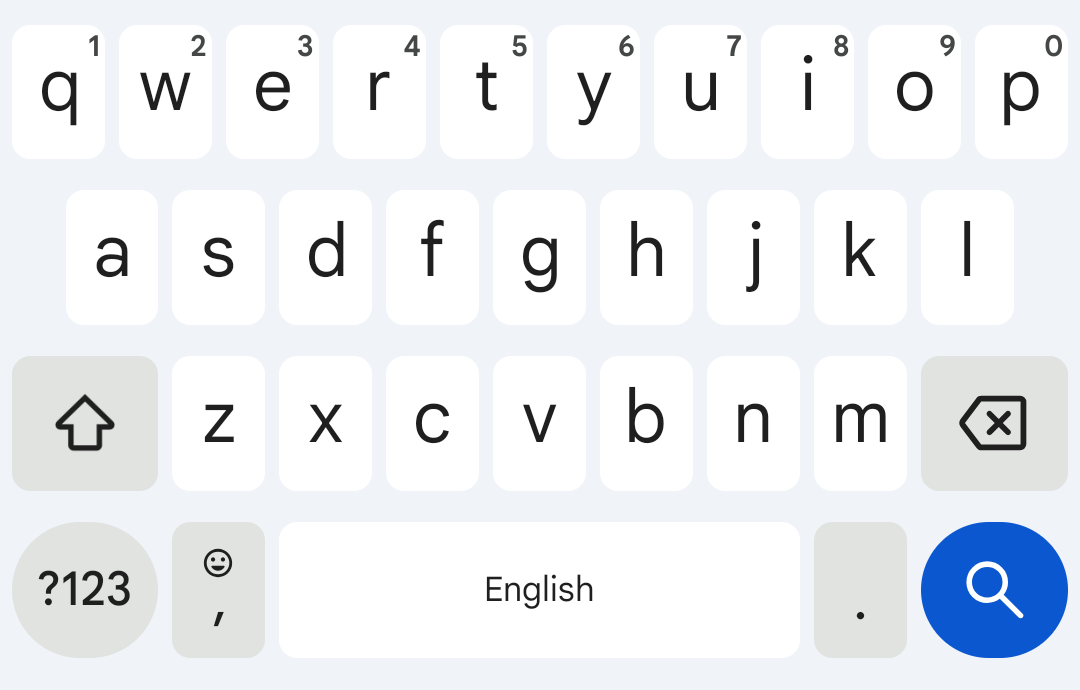}
        \caption{lowercase layout.}
        \label{fig:tracking}
    \end{subfigure}%
    \hfill
    \begin{subfigure}[b]{0.45\linewidth}
        \centering
        \includegraphics[width=\textwidth]{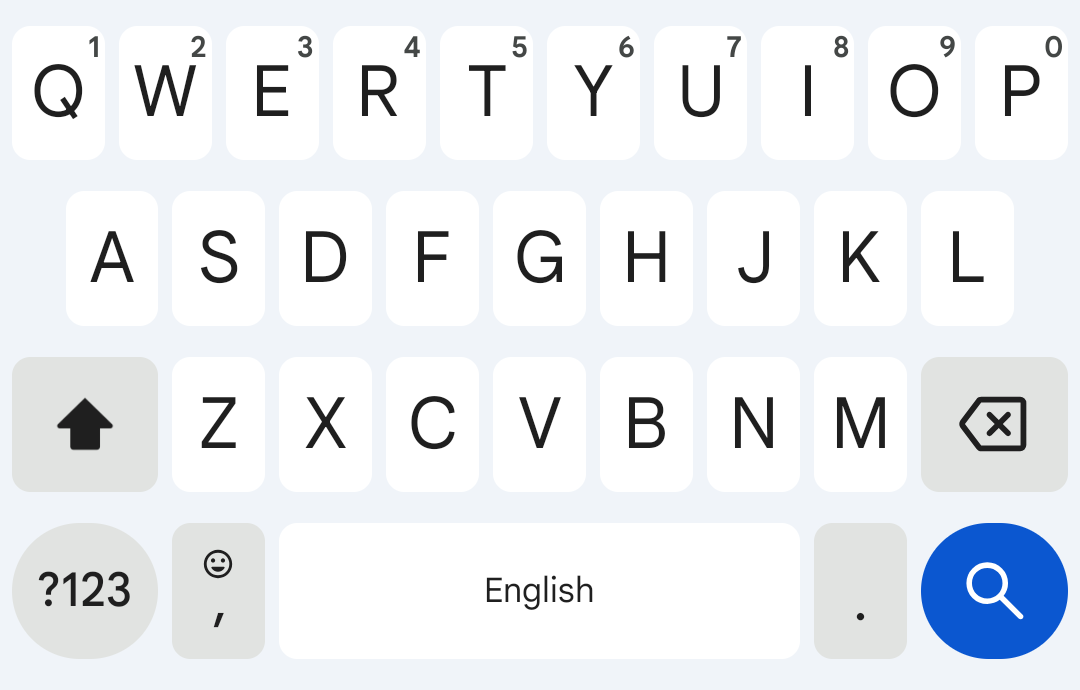}
        \caption{Uppercase layout.}
        \label{fig:detection}
    \end{subfigure}
     
    \vspace{0.2cm}
    
    \begin{subfigure}[b]{0.45\linewidth}
        \centering
        \includegraphics[width=\textwidth]{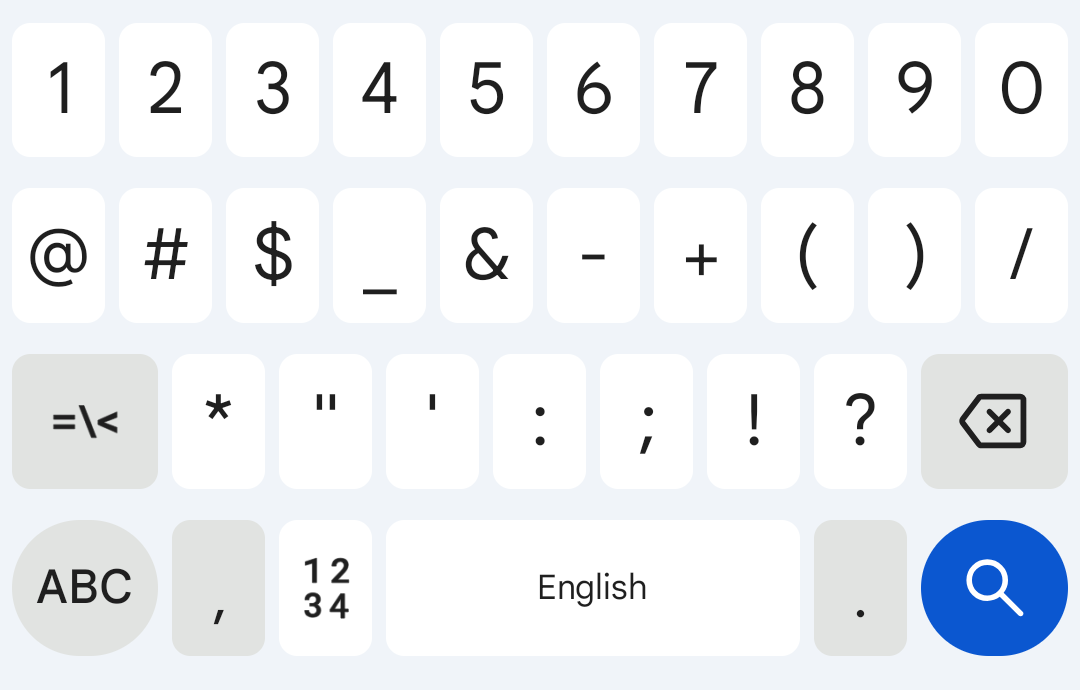}
        \caption{Numbers layout.}
        \label{fig:lm_correction}
    \end{subfigure}%
    \hfill
    \begin{subfigure}[b]{0.45\linewidth}
        \centering
        \includegraphics[width=\textwidth]{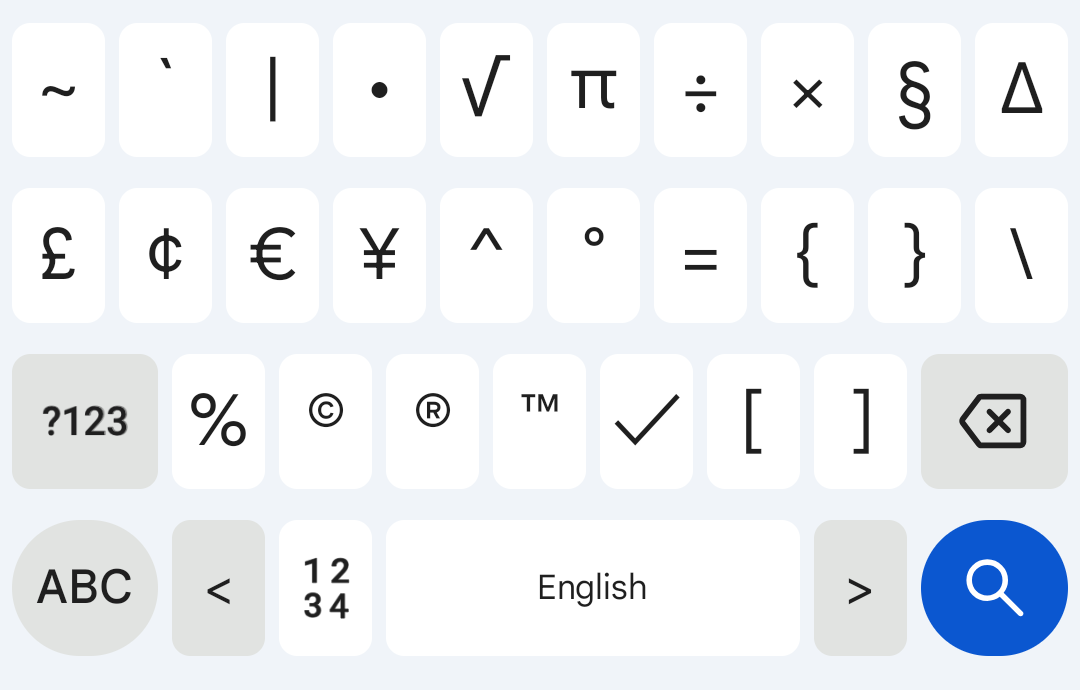}
        \caption{Symbols layout.}
        \label{fig:estimation}
    \end{subfigure}
    
    \caption{The four layouts of the Gboard (Google keyboard).}
    \label{fig:gboard_layouts}
\end{figure}

\begin{figure}[htbp]
    \centering
    \begin{subfigure}[b]{0.45\linewidth}
        \centering
        \includegraphics[width=\textwidth]{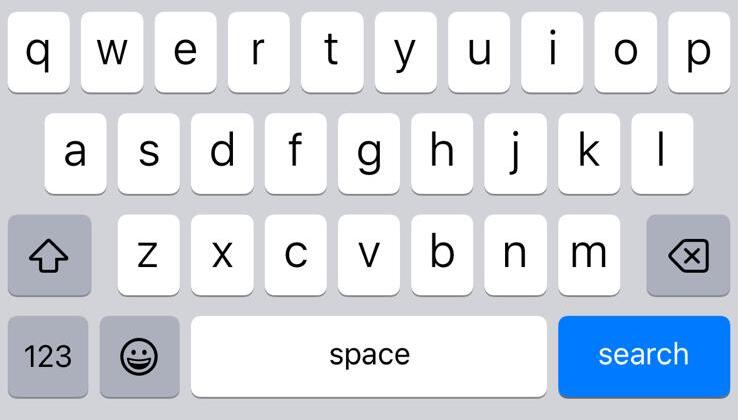}
        \caption{Lowercase.}
    \end{subfigure}%
    \hfill
    \begin{subfigure}[b]{0.45\linewidth}
        \centering
        \includegraphics[width=\textwidth]{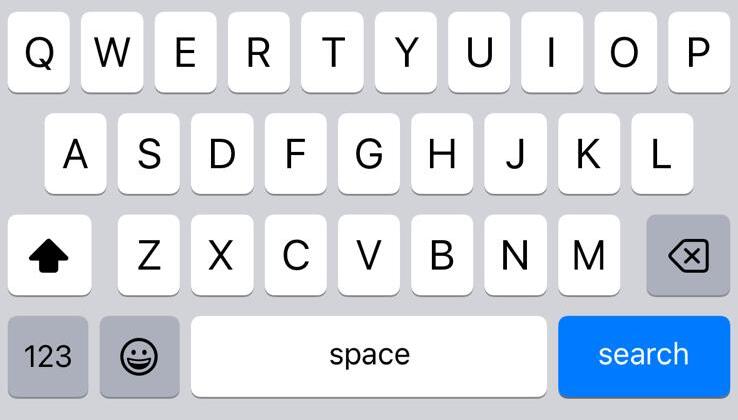}
        \caption{Uppercase.}
    \end{subfigure}

    \vspace{0.2cm}

    \begin{subfigure}[b]{0.45\linewidth}
        \centering
        \includegraphics[width=\textwidth]{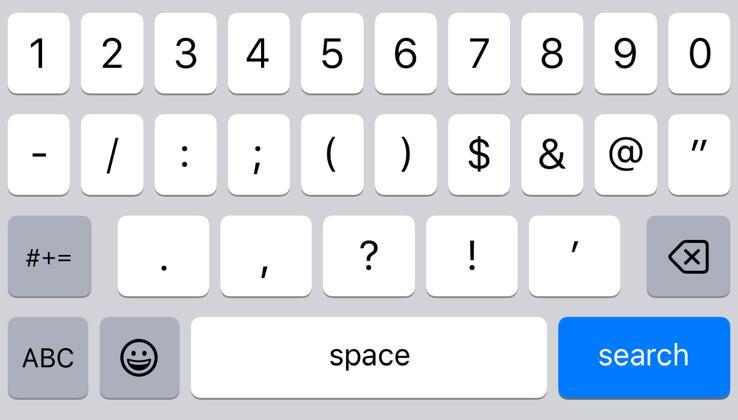}
        \caption{Numbers.}
    \end{subfigure}%
    \hfill
    \begin{subfigure}[b]{0.45\linewidth}
        \centering
        \includegraphics[width=\textwidth]{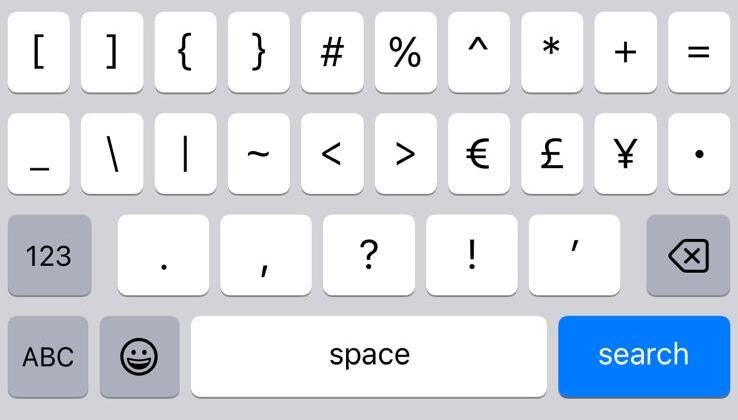}
        \caption{Symbols.}
    \end{subfigure}
    \caption{The four layouts of the iOS keyboard.}
    \label{fig:ios_layouts}
\end{figure}

\begin{figure}[htbp]
    \centering
    \begin{subfigure}[b]{0.45\linewidth}
        \centering
        \includegraphics[width=\textwidth]{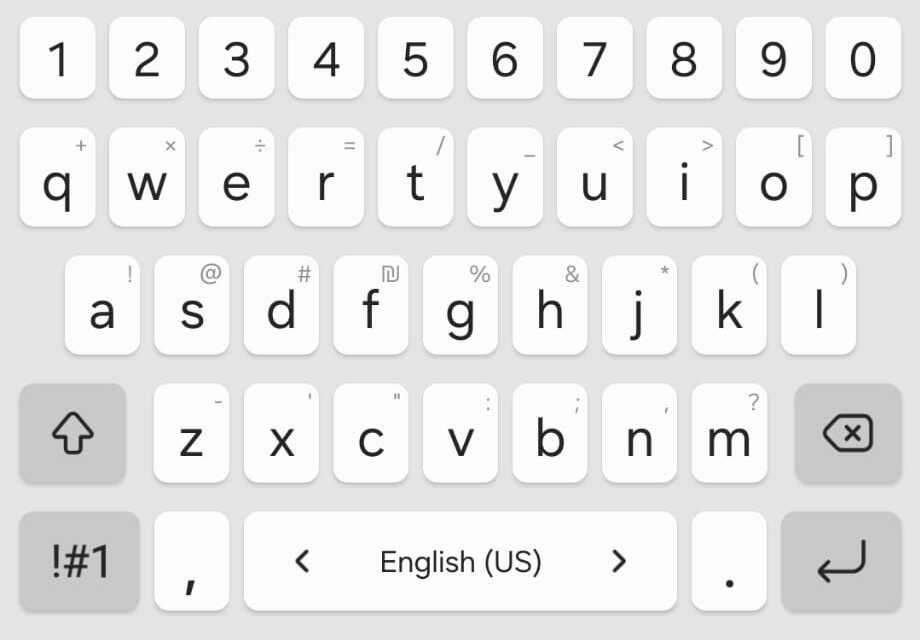}
        \caption{Lowercase.}
    \end{subfigure}%
    \hfill
    \begin{subfigure}[b]{0.45\linewidth}
        \centering
        \includegraphics[width=\textwidth]{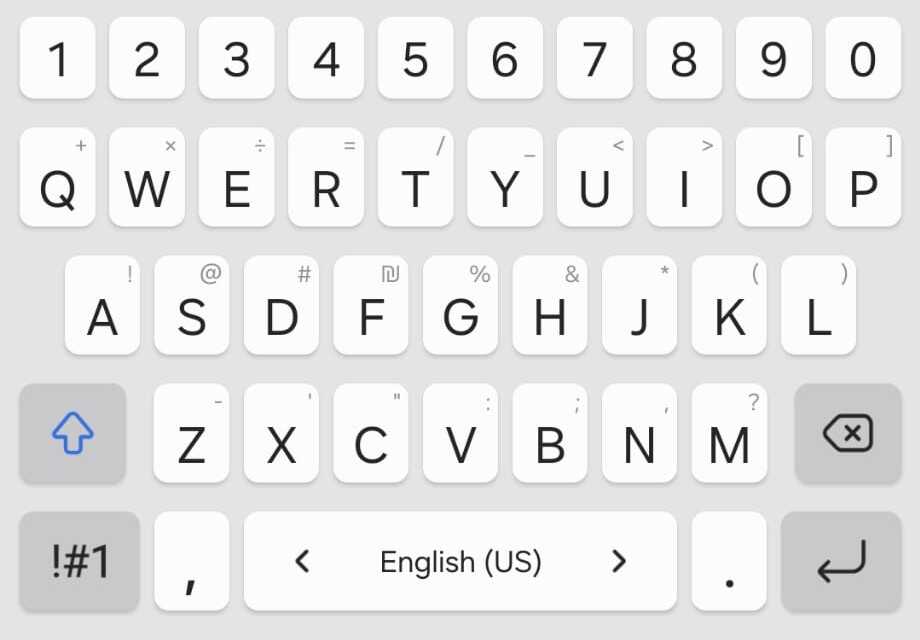}
        \caption{Uppercase.}
    \end{subfigure}

    \vspace{0.2cm}

    \begin{subfigure}[b]{0.45\linewidth}
        \centering
        \includegraphics[width=\textwidth]{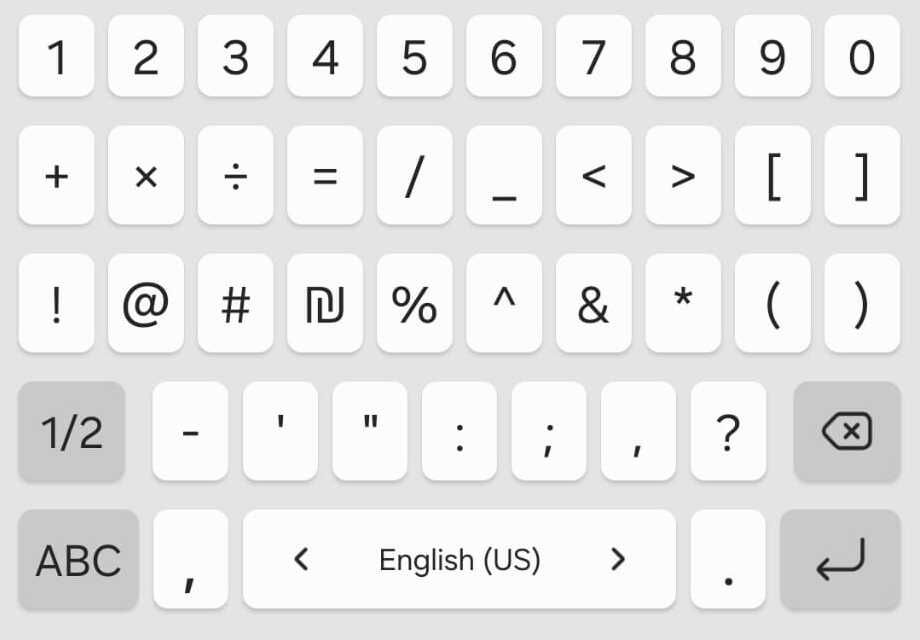}
        \caption{Numbers.}
    \end{subfigure}%
    \hfill
    \begin{subfigure}[b]{0.45\linewidth}
        \centering
        \includegraphics[width=\textwidth]{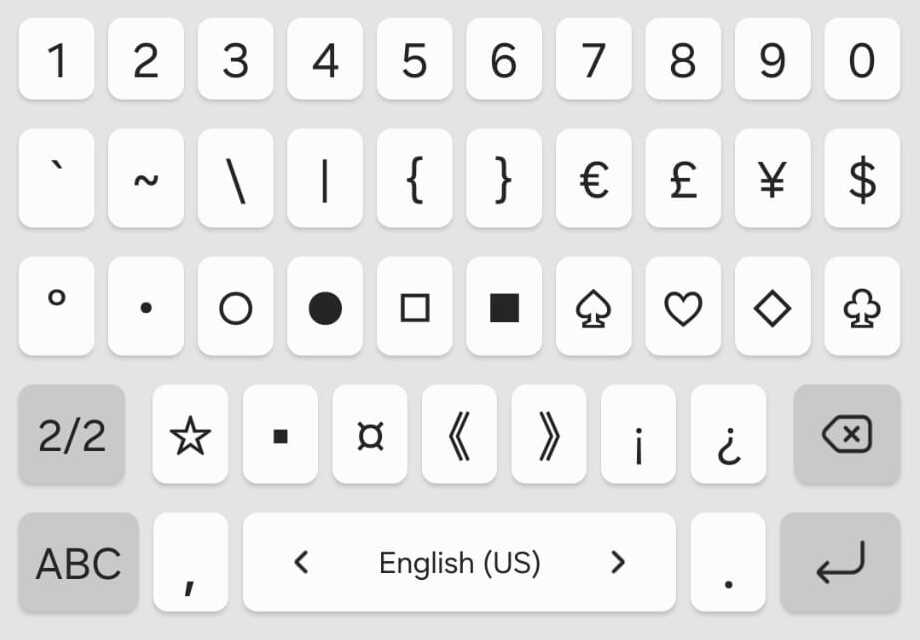}
        \caption{Symbols.}
    \end{subfigure}
    \caption{The four layouts of the Samsung keyboard.}
    \label{fig:samsung_layouts}
\end{figure}

Figures~\ref{fig:gboard_layouts}, \ref{fig:ios_layouts}, and \ref{fig:samsung_layouts} show the four layouts of each keyboard used in our evaluation. The keyboards differ in how characters are distributed across layouts and in the number of rows on the primary layout, so the same password can require a different number of touches and a different sequence of layout switches on each. This is the heterogeneity that the multi-configuration procedure of \S\ref{sec:step6} resolves.

\newpage
\section{Device and Software Versions}\label{app:devices}

Table~\ref{tab:devices} lists the smartphones used in our evaluation together with the operating-system and keyboard-application versions installed at the time of the study. All keyboards were used in their default configuration.

\begin{table}[htbp]
    \centering
    \small
    \begin{tabular}{@{}llll@{}}
        \toprule
        \textbf{smartphone} & \textbf{Screen} & \textbf{OS version} & \textbf{Keyboard (version)} \\
        \midrule
        Google Pixel~10   & 6.3\,in & Android \textsc{16} & Gboard \textsc{17.7.4} \\
        Apple iPhone~16   & 6.1\,in & iOS \textsc{26}     & iOS keyboard (built-in) \\
        Samsung Galaxy~A53 & 6.5\,in & Android \textsc{16} & Samsung Keyboard \textsc{5.9.20} \\
        \bottomrule
    \end{tabular}
    \caption{smartphones, operating systems, and keyboard applications used in the user study.}
    \label{tab:devices}
\end{table}

\newpage 
\section{Participants and Passwords}\label{app:participants_passwords}
\begin{table}[htbp]
    \centering

    \begin{tabular}{llccc}
        \toprule
        \textbf{Participant} & \textbf{Fingers used} & \textbf{Number of hands} & \textbf{Age} & \textbf{Gender} \\
        \midrule
        P0  & 2 Thumbs & 2 & 28 & M \\
        P1  & Index finger & 1 & 61 & F \\
        P2  & 1 Thumb & 1 & 37 & M \\
        P3  & 2 Thumbs & 2 & 39 & M \\
        P4  & 2 Thumbs & 2 & 37 & F \\
        P5  & 2 Thumbs & 2 & 35 & M \\
        P6  & 2 Thumbs & 2 & 34 & F \\
        P7  & 2 Thumbs & 2 & 31 & M \\
        P8  & Index finger& 1 & 28 & M \\
        P9  & 2 Thumbs & 2 & 29 & M \\
        P10 & 2 Thumbs & 2 & 24 & F \\
        P11 & 2 Thumbs & 2 & 21 & M \\
        P12 & Index finger& 1 & 51 & F \\
        P13 & 2 Thumbs & 2 & 18 & F \\
        P14 & Index finger& 1 & 56 & M \\
        P15 & Index finger& 1 & 65 & M \\
        \bottomrule
    \end{tabular}
    \caption{Typing behaviors and general information of our 16 participants (P0-P15)}\label{tab:paricipants_data}

    \vspace{2em}

    \begin{tabular}{cll}
        \toprule
        \textbf{Password length} & \textbf{Human-chosen} & \textbf{Password-manager-chosen} \\
        \midrule
        8  & Myspace0 & 9pkE!Bsy \\
        9  & Free2play & bSYjpP6i* \\
        10 & Junglelov3 & @2tfb6XjhR \\
        11 & Supording00 & yA\#4a*SaZUx \\
        12 & Apocaliptic4 & Py\&\$DGa3qWb6 \\
        13 & MatrixGamer07 & pZa\$r59nPaBS5 \\
        14 & 1Word1password & 6Y*ttY9vybK8Pf \\
        15 & Romeo\&Juliet101 & 7p5whefTpJ5*6\&b \\
        16 & \#100Mildoller\$!! & iLaE\%!H\&2wnoAtA! \\
        \bottomrule
    \end{tabular}
    \caption{Password test set for human-chosen and password-manager-chosen random passwords}\label{tab:password_test_set}
\end{table}

\end{document}